\documentclass[12pt,preprint]{aastex} 
\begin{document} 
 
\shorttitle{Debris disks in Pleiades} 
\shortauthors{Gorlova et al.} 
 
%________________________________________________________________ 
 
\title{\textit{Spitzer} 24 micron Survey of Debris Disks in the Pleiades} 
 
 
\author{Nadya Gorlova$^1$, George H. Rieke$^1$, James Muzerolle$^1$, John R. Stauffer$^2$, Nick Siegler$^1$,
Erick T. Young$^1$, \& John H. Stansberry$^1$}

\affil{$^1$Steward Observatory, University of Arizona, 933 North Cherry Avenue, Room N204, Tucson, AZ 85721-0065} \email{ngorlova@as.arizona.edu} 
\affil{$^2$Spitzer Science Center, California Institute of Technology, MS 314-6, Pasadena, CA 91125}

%% identified by the \altaffilmark after each name.  Specify alternate 

 
\begin{abstract}

We performed a 24 $\mu$m $2\degr \times 1\degr\,$ survey of
the Pleiades cluster, using the MIPS instrument on
\textit{Spitzer}. Fifty four members ranging in spectral type
from B8 to K6 show 24 $\mu$m fluxes consistent with bare photospheres.
All Be stars show excesses attributed
to free-free emission in their gaseous envelopes.
Five early-type stars and four solar-type stars
show excesses indicative of debris disks.
We find a debris disk fraction of 25\%
for B-A members and 10\% for F-K3 ones.
These fractions appear intermediate between those for younger 
clusters and for the older field stars.
They indicate a decay with age of the
frequency of the dust-production events inside the planetary 
zone, with similar time scales for solar-mass stars as have
been found previously for A-stars. 

\end{abstract}
 
 
%% subject header.  Objects should be in the appropriate "individual" 
%% headers (e.g. quasars: individual, stars: individual, etc.) with the 
%% additional provision that the total number of headers, including each 
%% individual object, not exceed six.  The \objectname{} macro, and its 
%% alias \object{}, is used to mark each object.  The macro takes the object 
%% name as its primary argument.  This name will appear in the paper 
%% and serve as the link's anchor in the electronic edition if the name 
%% is recognized by the data centers.  The macro also takes an optional 
%% argument in parentheses in cases where the data center identification 
%% differs from what is to be printed in the paper. 
 

%% \keywords{globular clusters: general --- 
%% globular clusters: individual(\objectname{NGC 6397}, 
%% \object{NGC 6624}, \objectname[M 15]{NGC 7078}, 
%% \object[Cl 1938-341]{Terzan 8})} 
 
 
\section{Introduction} 

After a few hundred million years, a planetary system is expected to have assumed 
its final configuration and has either set the stage for life, or will 
probably remain barren forever. It is difficult to probe this era. Most of its
traces have been obliterated in the Solar System. Only a minority of the
nearby stars are so young. Even for them, planets -- and particularly those
in the terrestrial planet/asteroidal region -- are faint and are lost in
the glare of their central stars. However, when bodies in this zone collide,
they initiate cascades of further collisions among the debris and between
it and other members of the system, eventually grinding a significant
amount of material into dust grains distributed in a so-called debris disk. Because 
the grains have larger surface area per unit mass compared to 
larger bodies, they (re)radiate more energy and therefore are 
more easily detected in the IR compared
to their parent bodies.
By studying this signal, we can probe the evolution of other
planetary systems through this early, critical stage.
 
Debris disks are found around stars generally older than $\sim$10 Myr, with no 
signs of gas accretion (as judged from the absence 
of emission lines or UV excess) \citep{Lagrange00,Hillenbrand05}. 
In the absence of gas drag, a 10 micron-size 
dust grain from the primordial, protoplanetary nebula cannot survive 
longer than $\sim$ 1 Myr within  10 $AU$ 
of a star due to a number of clearing 
processes, like sublimation, radiation pressure, Poynting-Robertson 
and stellar wind drag \citep{Backman93,Chen05a}. 
Therefore, any main sequence star older than 10 Myr with an infrared excess 
is a candidate to have circumstellar material supplied through 
debris disk processes.  
 
In our Solar System there are two main sources of dust: 
colliding asteroids and evaporating comets. 
The dust (grains $<$ 1 mm in size) is concentrated in two belts -- the asteroid belt 
at 2-3.5 $AU$, $L_{dust}=10^{-7}L_{\sun}, T_{dust} = 150-200$ $K, 
M_{dust} =10^{-8}M_{\earth} $ \citep{Low84,Dermott02,Hahn02}; 
and the Kuiper belt at 30-50 $AU$, $L_{dust}=10^{-7}L_{\sun}, T_{dust} =40-100$ $K, 
M_{dust} = 10^{-5} M_{\earth}$ \citep{Backman95,Morbidelli03}. 
Asteroid collisions that produce warm dust are driven by perturbations
by planets, a process that has cleared most of the interplanetary material
from the inner 30 $AU$ of the system.
In contrast, there is substantial mass in the Kuiper Belt,
concentrated in reasonably stable structures controlled by giant planets \citep{Liou99, MoroMartin02}.
Similarly, rings, blobs, warps and other non-uniformities are frequently observed in  
the spatially resolved images of nearby extrasolar debris disks
and inferred from the spectral energy distributions 
in the distant unresolved disks,
implying the presence of \textit{planetary mass bodies}
in these systems.
 
The first debris disks were discovered by \textit{IRAS} in the 1980s. 
Only a handful of nearby ones have been resolved in optical and near-IR scattered light 
or detected in submm thermal emission; \textit{ISO} has extended the \textit{IRAS}
results, but only to a modest extent \citep[e.g.,][]{Laureijs02}. The limited sensitivity
of \textit{IRAS} and \textit{ISO} and the even more severe limitations of other methods have
prevented a complete census, particularly to photospheric emission levels. 
In addition, some alleged disks have been confused with background objects or cirrus in the large beam 
of \textit{IRAS}.
The attempt to assess the percentage of stars 
with debris-disk emission therefore appeared premature  
from pre-\textit{Spitzer} data \citep{Zuckerman01}. 
 
With superior sensitivity and a smaller beam size, 
\textit{Spitzer} is providing many new insights to debris disk behavior. 
For example, \citet{Rieke05} demonstrated 
both the overall decline of debris disks with age (first noted in \citet{Holland98,Spangler01}) 
and the large scatter of disk properties at any given age 
(as previously noted in \citet{Decin03}). 
By probing excesses to within 25\% of the photospheric emission, 
\citet{Rieke05} found a surprising number of non-excess 
stars even at ages as young as 10 - 20 Myr, 
implying very rapid clearing of the inner $10-60 AU$ region in these systems. 
 
The study of Rieke et al. was purposely confined to 
A stars, with masses $\sim 2.5 M_{\odot}$. 
It poses the question of whether the same patterns 
of disk evolution hold for less massive stars, 
especially in the \textit{solar} mass range -- spectral types FGK. 
The study of excesses in young solar analogs could 
corroborate lunar geological evidence of the bombardment history 
in the Solar System. Two major surveys of solar type stars in the field 
are under way with \textit{Spitzer} -- the solar star survey by the MIPS GTO team, 
searching for disks around $\sim$150 field stars within 25 pc \citep{Bryden06}, 
and the Legacy survey by the FEPS group, searching for disks 
around $\sim$330 stars, both field and members of nearby open 
clusters and associations \citep{Meyer04}. 
Interpretation of these surveys is still in progress; nevertheless, 
preliminary results indicate that among solar analogs typically older than 1 Gyr 
$\sim$10-15\% have excess emission at 70 $\mu$m, 
indicative of Kuiper Belt analogs 
$\sim$10-100 times more massive than in the Solar System \citep{Kim05,Bryden06}.  
On the other hand, 24 $\mu$m excesses that 
would be indicative of massive zodiacal belts are 
much more rare around these stars -- $\sim$1\% \citep{Beichman05b,Beichman06}. 
 
To explore younger systems with good number statistics and small uncertainties in age determination, 
an open cluster survey has been initiated by the MIPS and IRAC GTO groups. 
The first clusters studied in this program were the 25-35 Myr old NGC 2547 \citep{Young04} 
and the 60-100 Myr old M47 (NGC 2422) \citep{Gorlova04}. 
At 24 $\mu$m photospheres were detected in NGC 2547 down to mid-F 
and in M 47 to early G spectral types, beginning to probe disks 
in the solar-mass regime. A decrease with age in both the amount and 
the fraction of excess (from $\sim$40\% to 25\%) is indicated for B-A stars 
between the two clusters (as included in Rieke et al. 2005). 
For F-K stars a similar trend is only suggested: 
quantitative estimates are impossible because photospheres 
start dropping below the detection limit. Thus, it is not clear whether
the high incidence of 24 $\mu$m excesses around A stars is a product
of their high luminosities and efficiency in heating dust, or
simply results because they are by definition relatively young ($<$ 1 Gyr).  
 
To probe the evolution of debris disks around solar-mass stars, 
we have investigated the Pleiades. 
The Pleiades is one of the best studied clusters 
with membership extending into the substellar  
regime \citep[e.g.,][]{Moraux03,Pinfield03}. 
At 24 $\mu$m {\it Spitzer} is potentially sensitive enough to detect photospheres 
in this cluster down to early M spectral type. 
To exploit these advantages, one must contend with 
an interstellar cloud of gas and dust, 
whose thermal emission beyond 10 $\mu$m 
dominates over starlight in the South-West 
part of the cluster. 
There is no consistent picture of the structure 
and kinematics of this nebulosity, 
except that it might be a part of the $\sim$10 Myr Taurus-Auriga 
starforming complex through or behind which the 
Pleiades happens to be passing at the present epoch \citep[e.g.,][]{Breger87}. 
Recent studies however agree 
that at least some of the material is located within 
the cluster and possibly interacts with the brightest 
members \citep{Gibson03,White03}. 
Dust in the molecular cloud is sheared by the radiation pressure and
assembles into filaments that wrap around stars, forming a characteristic pattern
of reflection nebulosity known as the "Pleiades phenomenon" \citep{Arny77,Herbig01}.
The unresolved thermal emission from the interstellar dust shell can mimic
emission from a debris disk, as was for example demonstrated with coronagraphic imaging of the 
\textit{IRAS-}selected Vega-like stars by \citet{Kalas02}.
Special measures were undertaken in our survey to avoid confusion with interstellar dust (\S\ref{red},\ref{sample}).

 Pre-\textit{Spitzer} IR and submm \citep{Zuckerman93} searches for Pleiades 
disks were largely unsuccessful. 
The few reported detections 
by \textit{IRAS} \citep{Castelaz87} and \textit{MSX} \citep{Kraemer03} 
concern the brightest B type members, 
which we find to be either surrounded by heavy cirrus or emission line stars 
with gaseous disks. 
\citet{Spangler01} observed 14 low mass members with \textit{ISO}, 
and claim detection (and excesses) at 60 and/or 90 $\mu$m 
in the two G0 and K2 members HII 1132 and HII 3163. 
The fractional dust luminosity in these stars, if accurate, 
is unusually high -- about $10^{-2}$, 
comparable to the transitional disks found around $\lesssim$10 Myr-old stars. 
Unfortunately both stars are outside of our \textit{Spitzer} fields. 
Recently \citet{Stauffer05} within the framework of the FEPS project 
investigated 19 solar-type members outside of the cluster core. 
They found excess emission at 24 $\mu$m at levels up to 40\% above the photosphere 
in a few members. 
Compared to the FEPS study, ours is centered on the cluster core, 
covers a larger area, and considers all members independent of spectral type 
or binarity. It is a magnitude- and background-limited survey 
based on the membership derived from the optical studies. For the
first time, it provides adequate number statistics to constrain the
disk incidence around Pleiades members similar in mass to the Sun. 

\section{Observations and Data Reduction}\label{red} 
 
The Pleiades were imaged with the MIPS instrument
in the scan map mode with half-pixel subsampling \citep{Rieke04}.
One area of size $90\arcmin\, \times 60\arcmin\,$ was centered on the cluster core,
as defined by the "seven sisters" (the brightest B-type members) and the 
associated reflection nebula. It was observed in two segments in February 2004. 
Another field of size $75\arcmin\, \times 45\arcmin\,$,
centered approximately $75\arcmin\,$ North of the cluster core, was observed in September
2004.  
All the observations were done using medium scan mode with half-array
cross-scan overlap, resulting in a total effective exposure time per point
at 24 $\mu$m of 80 seconds.  The 24 $\mu$m images were processed using
the MIPS instrument team Data Analysis Tool, which calibrates
the data and applies a distortion correction to each individual exposure \citep{Gordon05}.
We further applied a column-dependent median subtraction
routine to remove a small residual dark latent pattern from individual images
before combining into a final mosaic.
Data at 70 and 160 $\mu$m were taken
simultaneously; however, the images yielded no detections of interest
to this study because of the sensitivity limits and strong background cirrus,
and we will not discuss them further.

\begin{figure} 
\plotone{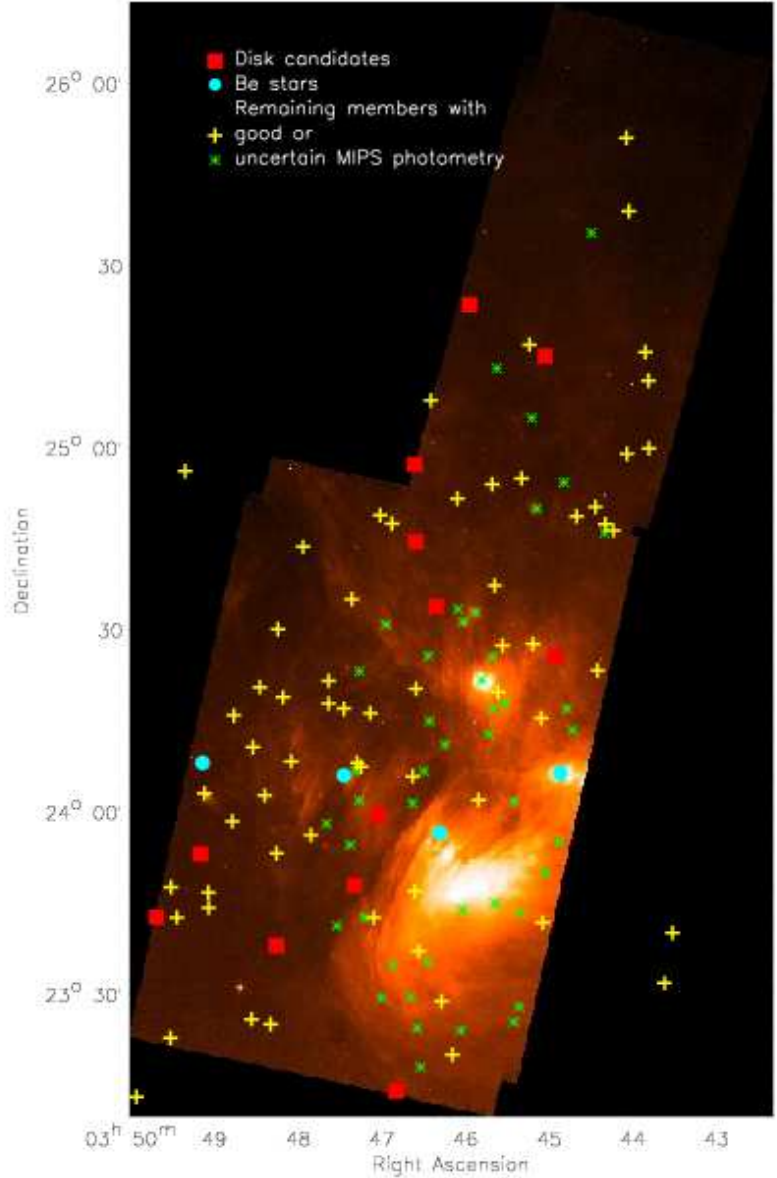}   
\caption{MIPS 24 $\mu$m mosaic made of three scan maps. 
\textit{Crosses:} Pleiades members with 24 $\mu$m images allowing 
accurate PSF-fitting photometry, from this work and \citet{Stauffer05}, 
also listed in Table \ref{table2}. 
\textit{Squares:} debris disk candidates, highlighted in bold in Table \ref{table2}. 
\textit{Asterisks:} Pleiades members with 24 $\mu$m images contaminated by cirrus or 
a nearby source, listed in Table \ref{table1}. 
\textit{Circles:} Be stars, all of them formally showing excess, a mix of point-like and  
extended sources. 
}\label{fig1}   
\end{figure}

The thermal emission from the nebula is strongest in the West side of the cluster,
where the ISM cloud is of highest density as indicated by the
enhanced extinction, polarization and detection of CO gas 
\citep{Breger87}. This highly non-uniform nebulosity limits our ability
to perform photometry on the faint sources in the central field
to only those that are situated
in the clean gaps between the the bright filaments.
The Northern field contains fewer members, being further from the
cluster core, but it has the most uniform background. 
We performed photometry independently on each field,
which allowed us to evaluate the uncertainty of our photometry
based on the objects from the overlapping regions $\sim5\arcmin\,$-wide.
The full area of our survey is shown in Fig. \ref{fig1}.
The image was created by merging all three fields
with the \textsc{iraf} task  \texttt{imcombine}, after applying appropriate offsets 
to account for the different levels of zodiacal light in each field.
Stars marked in the figure outside the map boundaries are from \citet{Stauffer05}.

The complicated background required an investigation 
of the optimum way of doing photometry. 
We carried out two tests to find parameters  
that provide the most accurate results 
for aperture and PSF-fitting 
photometry. 
Standard \textsc{iraf} tasks \texttt{phot} and  
\texttt{allstar} from the \texttt{daophot} package were used.  
The first test compared the difference between the input 
and the recovered flux of artificial stars, inserted into 
different background environments (using task \texttt{addstar}). 
The second test looked at the tightness of the locus of the cluster members 
on the J-H vs. K-[24] diagram. For the non-excess stars, we assumed 
that the dispersion is dominated by the measurement errors in the 24 $\mu$m flux, 
since reddening is negligible at these wavelengths. 
We tried a range of aperture radii starting with the HWHM of the MIPS beam  
of 2.3 interpolated pixels (1 pixel $= 1.247$\arcsec\,), 
and a range of sky annuli centered at the first Airy ring at 7 pixels. 
We finally adopted PSF-fitting photometry with a small value for the fitting radius of 3 pixels 
and a close sky annulus with inner and outer radii of 5 and 10 pixels respectively. 
The PSF template for all three fields 
was constructed from 9 stars in the North field, 
carefully selected to be in the regions clear of nebulosity
and to have a representative range of magnitudes.  
As a check, another PSF was created from 7 'clean' stars eastward from the cluster center. 
Magnitudes of the PSF stars measured with both PSFs agree to within 
0.01 mag. 
 
To determine an aperture correction, 
we compared magnitudes within an aperture radius of 3.5 pixels 
for the 9 PSF stars performed with our adopted \textit{PSF-fitting} photometry 
to the \textit{aperture} photometry with parameters 
that would normally be used on a uniform background --
aperture radius equal to the FWHM of the beam (4.6 pixels), sky annulus beyond the 
first Airy ring (12 and 18 pixels for the inner and outer radii),  
and aperture correction of 0.68 mag from 4.6 pixels to infinity  
(as established by the MIPS team based on the STinyTim simulated PSF).
The resulting aperture correction for our Pleiades-customized PSF-fitting photometry 
is 0.89 mag.  
Finally, to translate fluxes in number counts into 24 $\mu$m magnitudes,   
we use a conversion factor of $1.627\cdot10^{-6}Jy/(DN/s)$ (where 
$DN$ is the Data Number per pixel, for 1.25\arcsec\, pixels). 
We took 7.3 Jy for the [24] magnitude zero point. 
 
Two cluster members (HII 1284 and HII 1380), of [24]$\approx$7$^{m}$ (or $\sim$ 10 mJy) 
appear on both East and West sub-fields, providing a consistency check
of the photometric errors calculated by \texttt{allstar}. Their magnitudes differ between the 
two sub-fields by 0.08 and -0.01 mags respectively (with the larger difference for HII 1284
most certainly due to a faint companion, \S \ref{Binarity}).
We also detect six members in common with \citet{Stauffer05} (HII 173, HII 174, 
HII 250, HII 314, HII 514, HII 2147). The average difference between our MIPS [24] magnitudes 
and Stauffer et al.'s is -0.02, RMS$=$0.10 mag. 
We have four early-type stars in common with \citet{Rieke05} 
(HII 1431, HII 1823, HII 1876, HII 2425); comparison 
with updated photometry (K. Y. L. Su, private commun.) 
indicates 0.07 mag difference between us and 
the latter work, with RMS$=$0.04 mag.
We conclude that the random error of our 24 $\mu$m photometry 
is about 5\% for objects brighter than $\sim$2 mJy in the clean regions of the sky, 
which is consistent with the scatter seen in the K-[24] color of the non-excess members 
(\S \ref{excess}). The systematic error in our data calibration
is of similar magnitude (few percent), consistent with comparisons of objects
common to other independent studies.
 
\section{Sample}\label{sample} 
 
We measured the 24 $\mu$m fluxes of the Pleiades members as follows. 
Using the \texttt{daophot} task \texttt{find} and visual inspection, 
for each MIPS field we constructed an initial coordinate list for all 
discernible point-like sources. 
Task \texttt{phot} provided initial aperture photometry, 
which served as input for the PSF-fitting photometry with \texttt{allstar}.  
Task \texttt{allstar} was executed with the recentering option turned on. 
Next, the \texttt{allstar} coordinates were correlated with the 2MASS point-source catalog, 
with a search radius of 2.5\arcsec\,. 
Using the 2MASS designations, we identified Pleiades members 
using an on-line catalog of the highly probable members maintained 
by J. Stauffer. Tables \ref{table1} and \ref{table2} summarize our measurements 
as well as some catalog properties for these sources.  
The information on binarity was supplemented with data 
from \citet{Liu91} and \citet{Raboud98}.
Table \ref{table2} also contains measurements from \citet{Stauffer05}
for 13 members outside of our field of view.
 
\begin{figure}  
\plotone{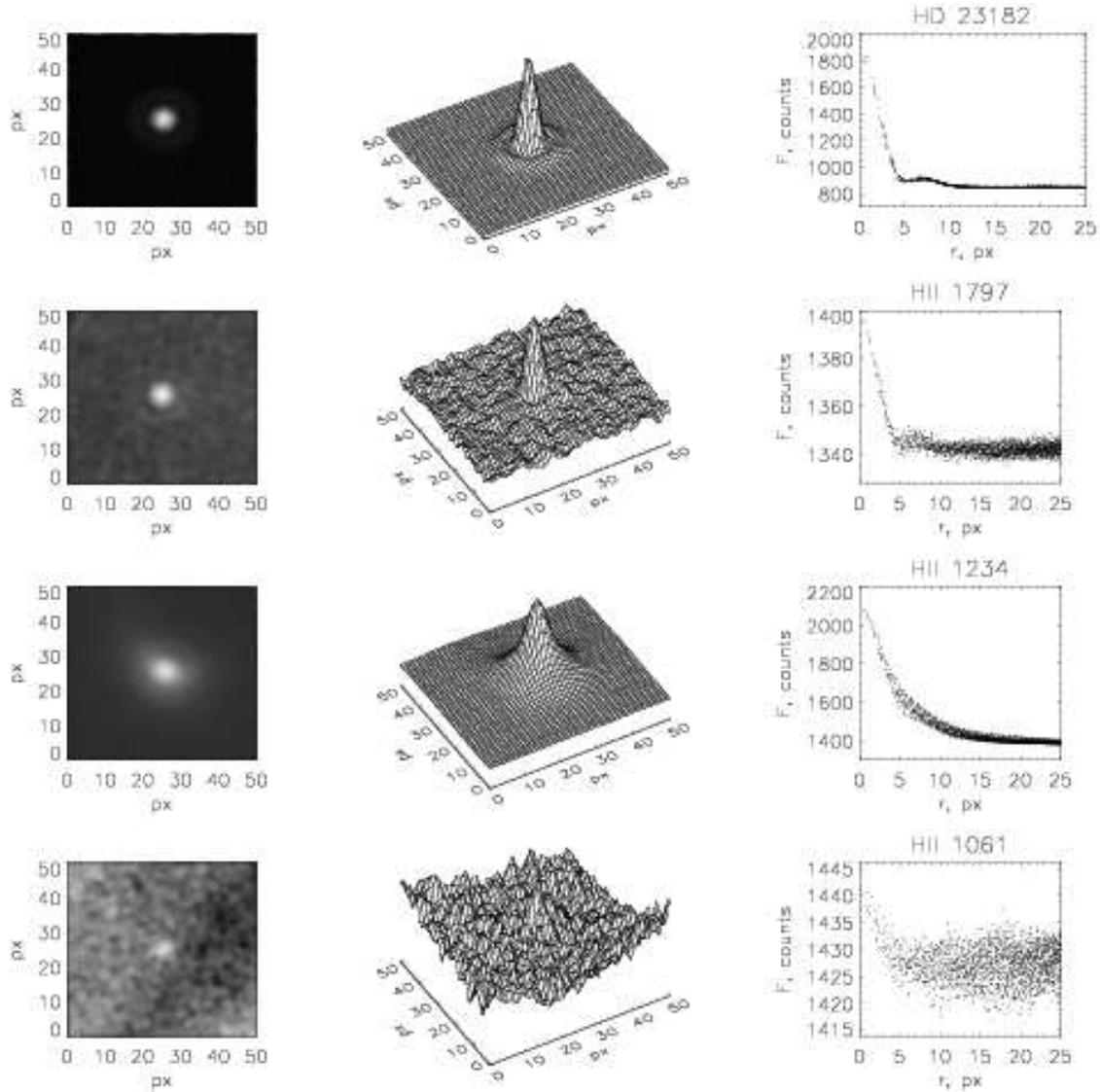}   
\caption{The surface and radial plots are shown for representative 
'clean' and contaminated 24 $\mu$m sources. 
HD 23182, a field star, is one of our PSF templates. 
HII 1797 is a Pleiades member from Table \ref{table2}, an example of the sources 
with well-defined, point-like profiles. 
HII 1234 and HII 1061 are from Table \ref{table1},  
examples of bright sources with extended smooth halos, 
and of faint sources with non-uniform background, respectively. 
}\label{fig2}   
\end{figure}

\begin{figure} 
\plotone{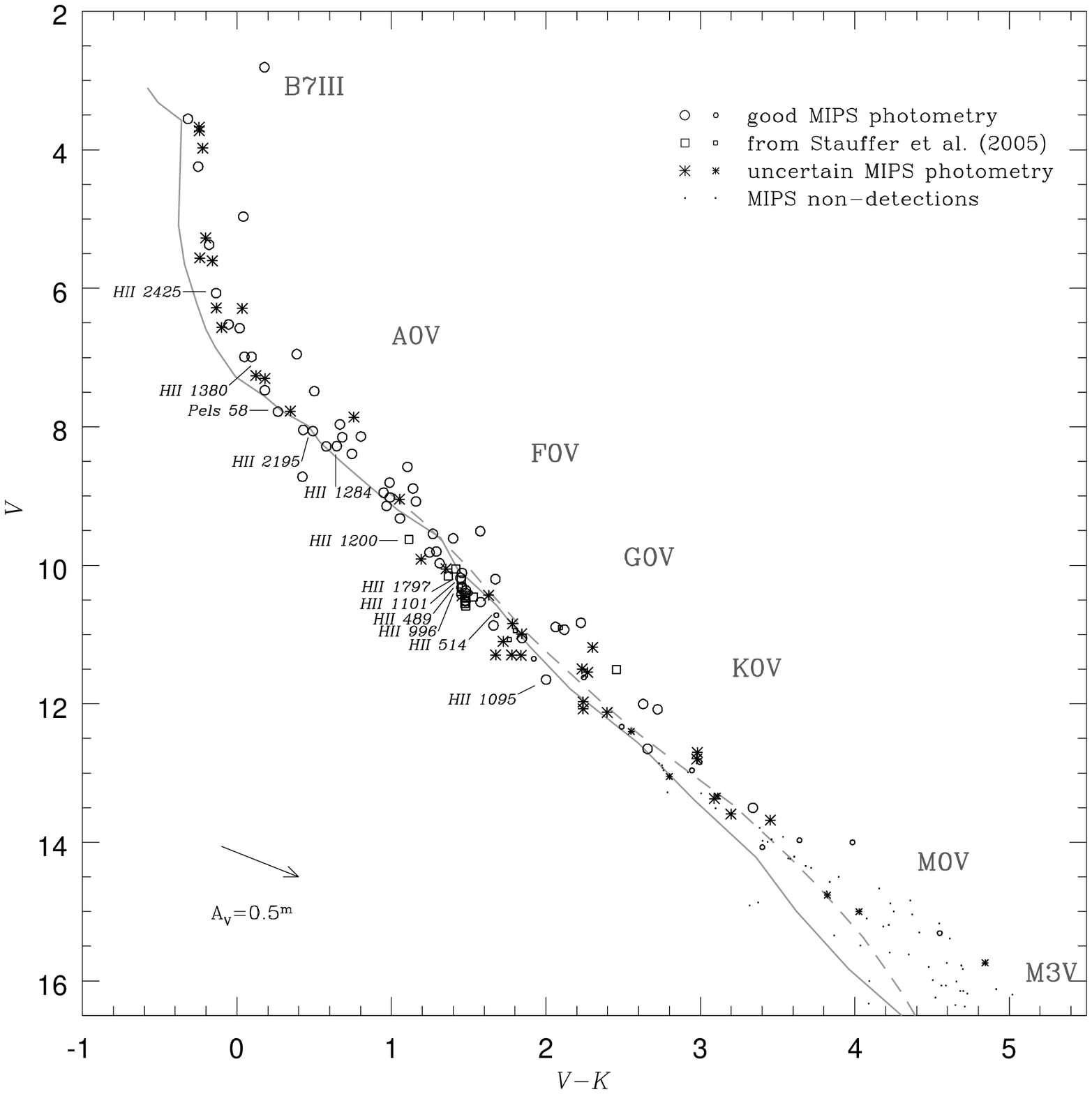}   
\caption{Pleiades color-magnitude diagram. \textit{Open symbols:} 'clean' 24 $\mu$m sources 
from Table \ref{table2}, circles -- this work, squares -- from \citet{Stauffer05}. 
\textit{Asterisks}: contaminated 24 $\mu$m sources from Table \ref{table1}.  
Big symbols: objects with spectral type, dereddened; 
small symbols: without spectral type, non-dereddened. 
\textit{Dots}: 24 $\mu$m non-detections. 
Marked are debris disk candidates according to this work or \citet{Stauffer05}. 
\textit{Solid line}: a 115 Myr isochrone 
by \citet{Siess00}. \textit{Dashed line}: a 125 Myr isochrone by \citet{Baraffe98}.
Both isochrones are of solar metallicity and assume a distance of 133.5 pc \citep{Soderblom05}. 
}\label{fig3}   
\end{figure}

Meaningful PSF photometry cannot be performed if the sky level is highly 
variable within the fitting region, or if the object's shape deviates  
significantly from an Airy function with the adopted FWHM. Also compact cirrus falling within the 
6\arcsec\, MIPS beam (corresponding to 800 $AU$ at the Pleiades distance of 133 pc) 
can mimic radiation from an unresolved debris disk.  
To exclude ambiguity in the photometry, 
we examined the image of every detected member. 
We designate as problematic objects having either  
a non-uniform (filamentary or curved) background within $\sim$ 15 pixels of   
the target (1.5 times the region of sky fitting) or a bright companion contaminating the target aperture. 
For the sources that passed this test, we generated radial profiles and
rejected any with FWHM $\gtrsim$ 5 pixels (6.2\arcsec). 
These problematic sources are listed in Table \ref{table1} and examples are shown in Fig. \ref{fig2}; 
in Figs. \ref{fig1} (except for Be stars) and \ref{fig3} they are designated by asterisks. 
Most of them cluster around the ridge of dense ISM cloud in the West part. 
For a number of them (highlighted in bold in Table \ref{table1}) 
the formal procedure of PSF-fitting indicates considerable excesses. 
Whether these excesses are real and due to debris disks or an artifact of cirrus contamination 
remains to be seen with better resolution. 
The analysis presented in the rest of the paper only deals 
with members that passed the above contamination test. 
These 'clean' sources are shown in Table \ref{table2} and in Fig. \ref{fig4}. 
 
As a measure of the 24 $\mu$m excess we use a $K-$[24] color 
(where $K$ is a 2MASS $K_{S}$).
Since there is no NIR excess, as expected for stars of age $\sim$100 Myr,
the $K$ band flux represents the photospheric flux.
$K$ band is the least affected by reddening of the three 2MASS bands. 
On average, the extinction toward the Pleiades is small, $<A_{V}> \approx$ 0.1 mag. 
However, it may exceed 1 mag in the South-West, the densest part of the 
interstellar cloud \citep{Breger86,Breger87,Stauffer87,Gibson03}, 
resulting in $\Delta K > 0.1$ mag. Considering 
that this is above the uncertainty 
of our photometry, red $K-[24]$ color due to obscuration may be misinterpreted as a weak 24 $\mu$m excess. 
Therefore, we corrected the $V$ and $K$ magnitudes for extinction. The 
extinction values reported in Tables \ref{table1}, \ref{table2} 
are either from the Pleiades catalog or have been derived 
by us for members with known spectral types. 
We calculated reddening using $V-K$ and $J-K$ intrinsic colors for dwarfs 
tabulated in \citet{Bessell88} (after 
converting them into the 2MASS system following transformations in \citet{Carpenter01}). 
The following reddening laws have been used: 
$A_{V}=3.1 E(B-V)=3.1 E(b-y)/0.74=A_{J}/0.282=A_{K}/0.112$ 
\citep{Cambresy02}. 
Extinction in the 24 $\mu$m band has been neglected. 
Big symbols on Figs. \ref{fig3}, \ref{fig4} represent dereddened magnitudes; 
small symbols are observed ones, since no spectral type information is available for these fainter members. 
 
Fig. \ref{fig3} shows a color-magnitude diagram for our and Stauffer's combined sample. 
The upper envelope traces equal mass binaries. 
The debris disk candidates (\S\ref{excess}) are marked; 
none of them appears to lie on the binary sequence. 
Non-detected members within the MIPS field of view are shown as dots. 
One can see that at the age and distance of the Pleiades 
we are 100\% complete at detecting 
photospheres earlier than K3 (0.8M$_{\odot}$), 
and in the absence of interstellar dust (``cirrus'') could 
potentially sample photospheres as
late as $\sim$M2 (0.4M$_{\odot}$). 
  
\section{Excess objects}\label{excess} 
 
\begin{figure} 
\plotone{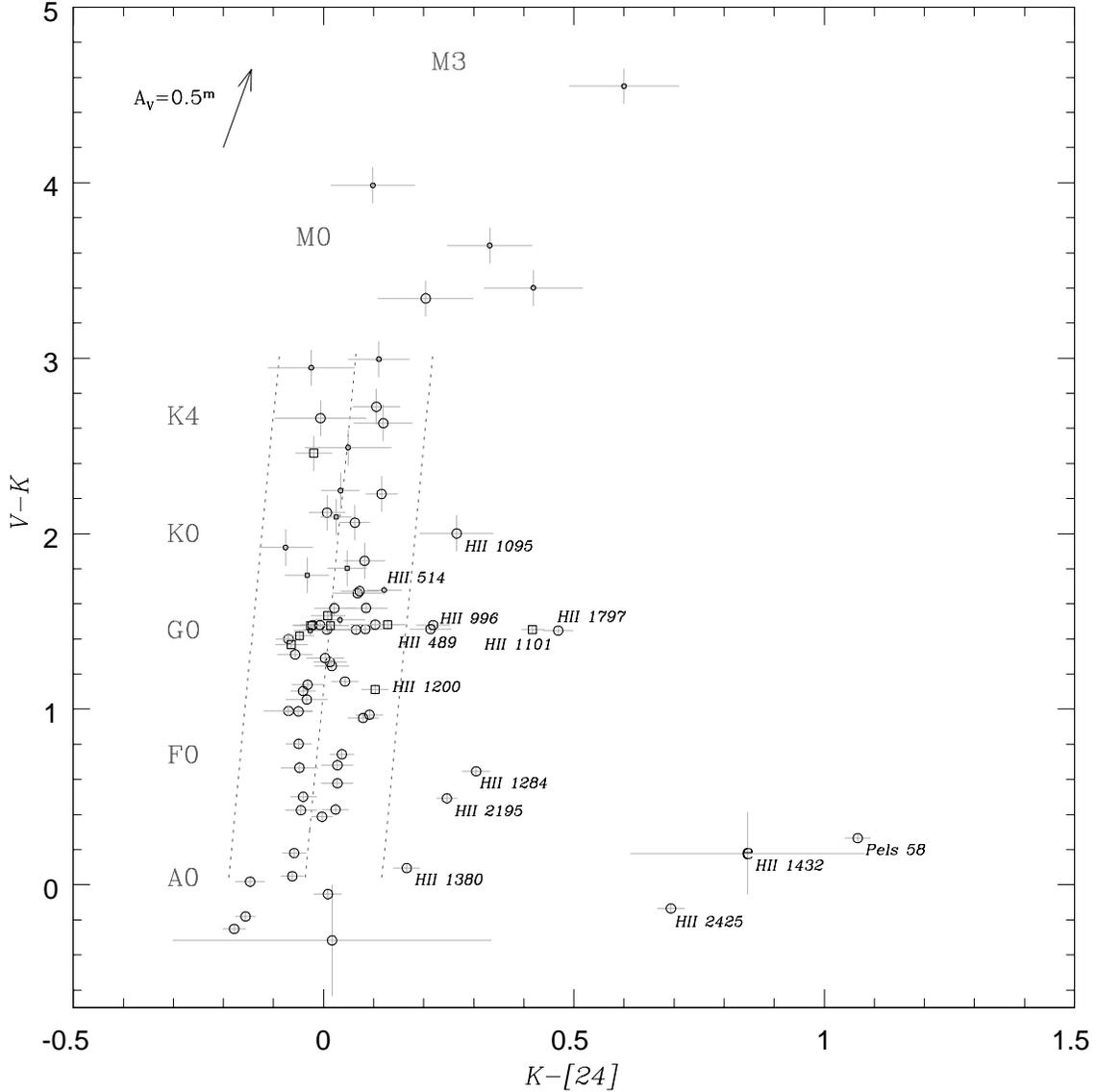}   
\caption{$V-K$ vs. $K-[24]$ diagram identifying excess objects (marked). 
Only members with good 
24 $\mu$m photometry (Table \ref{table2}) are plotted. 
Symbol coding is the same as in Fig. \ref{fig3}. 
``e''-Be star HII 1432; another Be star, 
HII 2181 (Pleione), is off scale at  $K-[24]=2.3$. 
$V$ magnitudes are a compilation from Stauffer's catalog, 
we estimate their errors to be 0.02 mag for B-G stars ($V<11^{m}$) 
and 0.1 mag for later types; errors in $[24]$ do not include the systematic error 
of about 5\% from uncertainty in aperture correction and  
in absolute calibration of the 24 $\mu$m array. 
\textit{Dotted lines:} linear fit to the non-excess A-K members and $\pm$ 3 standard deviations 
in $K-[24]$ from it.  
}\label{fig4}   
\end{figure}

Fig. \ref{fig4} shows the color-color diagram we  
use to identify members with excesses. 
Only objects with high quality 24 $\mu$m photometry are plotted. 
The $V-K$ axis represents photospheric colors. 
The $V$ and $K$ bands are separated far enough in wavelength space  
to trace temperature/spectral type and hence luminosity and mass for 
main sequence stars. 
The NIR colors alone, though less affected by reddening, 
provide a smaller magnitude range and become non-monotonic  
at the B-A and K-M transitions. 
The $K-[24]$ color on the other hand 
only weakly depends on the stellar temperature, 
since both bands fall on the Rayleigh-Jeans  
tail for objects as cold as early M dwarfs (T$_{eff}>$3200 K). 
In the 'Vega system' this color should therefore stay close to 0. 
Significant positive values indicate the presence 
of a circumstellar component. 
 
Indeed, as Fig. \ref{fig4} shows, the majority of stars between  
$V-K=$ 0.05 to 3 (spectral types A to K5) cluster around $K-[24]=0$, 
getting slightly redder with later spectral type. 
We regard this region as the locus of pure photospheric colors. 
It can be fitted as  
 \begin{eqnarray*} 
(K-[24])_{phot}=0.034(\pm 0.010)\times (V-K)_{phot}-0.038(\pm0.016) 
\end{eqnarray*}  
The standard deviation of residuals for the 57 fitted  
stars is $\sigma(K-[24])=0.051$ mag, which is within the random uncertainty 
of our photometry. 
 
Eleven stars however, including HII 1101 from \citet{Stauffer05}, 
lie significantly redward of the non-excess locus, 
indicating $K-[24]$ excess above the 3$\sigma$ detection level. 
HII 1095 is another possible excess member,
whose larger uncertainty puts it just below the 3$\sigma$ confidence level.
Two more, HII 1200 (outside of our field of view) and HII 514, 
may possess weaker excesses according to the photometry of \citet{Stauffer05}.
Despite the difference in the integration times
(400 s per source in \citet{Stauffer05} and 80 s in our survey),
both studies observe a similar scatter in the $K-[24]$ color for
the solar-type ``non-excess'' stars -- 0.04-0.05 mag from the average 0 value.
This makes it difficult to prove excesses in these two stars
beyond the 3$\sigma$ level; only HII 514 would pass the 0.15 mag threshold adopted in this work.
All 14 suspected excess candidates are marked in Fig. \ref{fig4} and highlighted in Table \ref{table2};
they (as well as HII 152) have been excluded from the above fit to non-excess members.
Excesses are seen around all 
spectral types down to K0 (and possibly mid-K, Table \ref{table1}), with values ranging 
from 0.1--0.5 mag for solar-type stars and 0.15-2.3 mag for early-type stars.   

Could excess emission arise from a nearby object
confused with a Pleiades member?
The average offset between MIPS and 2MASS positions
for members from Table \ref{table2} is 0.8\arcsec\,.
Except for HII 514 and HII 1095, which are the faintest excess candidates, 
the 24 $\mu$m positions fall within a r$=$0.5\arcsec\, circle 
around the 0.8\arcsec\, systematic offset on the $\Delta RA$ vs. $\Delta DEC$ plane,
together with the majority of the non-excess members.
Furthermore, they have flux densities above 2 mJy, and
the previous studies have shown that the probability of a spurious 
excess due to chance alignment with a background giant star or galaxy
at this flux level
is small -- $\lesssim$0.1\% per source \citep{Gorlova04,Stauffer05}. 
Due to the proximity of the Pleiades to the ecliptic however, 
there is concern about asteroid contamination. 
We examined images of the detected Pleiades members
for the presence of nearby companions
without a 2MASS counterpart -- signaling a medium-speed asteroid leaving
a chain of images as it moves between successive scans.
Also, when the object fell in the overlapping region between two fields,
we could identify contamination by fast moving asteroids by comparing
two images obtained a few days to a half year apart.
A 'close companion' to HII 320 from Table \ref{table1} is
one such example of likely asteroid contamination, but none are among the 
14 excess candidates.
 
The strongest excess among point-like sources belongs to 
HII 2181 (28 Tau, Pleione). The third strongest excess is for 
HII 1432 (25 Tau, $\eta$ Tau, Alcyone), the optically brightest member (Fig. \ref{fig3}). 
Both stars belong to the Be, or 'classical', type of 
emission-line B-A stars. 
Two other emission stars,  
HII 468 (17 Tau, Electra) and HII 980 (23 Tau, Merope), 
may also possess excesses of $\sim$0.5 mag, but 
unfortunately bright nebulosities surrounding these stars make our measurements 
uncertain (placing them in Table \ref{table1}). 
All four are post-MS stars, which is indicated 
by their giant or sub-giant luminosity classes, 
and fast rotators ($vsini > $200 km/s). 
The excess emission is probably free-free emission from 
gaseous Keplerian disks produced by radiation-driven mass loss and fast rotation
\citep{Waters86,Porter03}. 
The emission stage is episodic -- once in about $\sim$10 years 
for reasons still debated these stars eject a shell 
that settles into a temporal disk in the equatorial plane. 
In some cases  
shell episodes coincide with the periastron passage  
of a companion \citep{Gies90}. Mass transfer in a binary system could also account 
for the spin up of the primary \citep{Harmanec02}. 
Pleione is a good example -- fastest rotator, with biggest IR excess 
and a companion resolved with speckle interferometry \citep{McAlister89}. 
On the other hand, any companion to Merope has escaped detection so far \citep{Raboud98}; 
the only indirect evidence for one is the prominent X-ray emission \citep{Daniel02}.
 
In the absence of spectroscopic information for the photometrically selected 
members in clusters other than the Pleiades, debris disk stars can be distinguished from 
emission-line stars by the absence of excess NIR emission in the former, 
since even the hottest candidate debris disks have T$<900$ K \citep{Uzpen05}. 
Indeed, none of the rest of the 12 excess members possesses an excess in the 2MASS colors,
nor is mentioned as an emission-line star.
We therefore regard these stars as debris disk candidates.
 
\begin{figure} 
\plotone{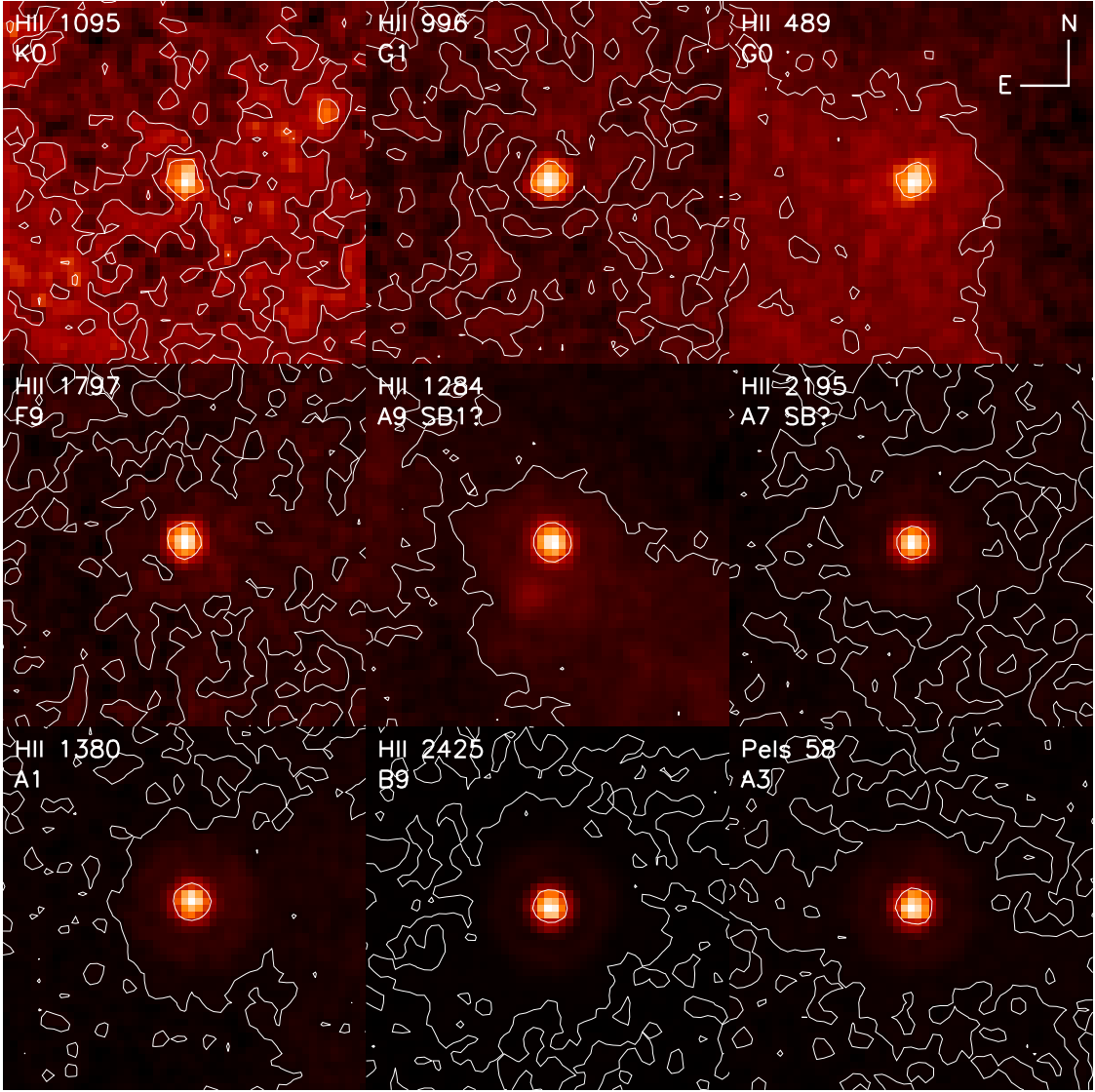} 
\caption{24 $\mu$m images of the debris disk candidates 
discovered in this work; $64\arcsec \times 64\arcsec$ (51 px $\times$ 51 px), linear flux scale. 
The central contour is drawn at the half-maximum of the source flux
and the remaining contours at the median sky level. 
}\label{fig5}   
\end{figure}

The 24 $\mu$m images of the 9 debris disk candidates discovered in this work
are shown in Fig. \ref{fig5};
the remaining 3 are from \citet{Stauffer05}.  
The central contour is drawn at 1/2 of the intensity of the brightest pixel on the source, 
as a measure of the FWHM. As can be seen from these images, 
the excess emission appears symmetrical and unresolved 
within the MIPS beam of 6\arcsec, meaning that it is confined to  
$<$800 $AU$ from the parent stars, consistent with the 
sizes of the nearby debris disks resolved in the mid-IR 
\citep{Stapelfeldt04,Su05,Telesco05}. 
The remaining contours are drawn at the level of the median sky intensity, 
representing the quality of the background. 
In all cases the background is flat or only tilted 
(HII 489 and HII 1284), but uniform within at least the first Airy ring 
where we measure the sky level, 
lending confidence to the reality of the detected excesses 
(compare with the uncertain excess candidates HII 1234 and HII 1061 
from Table \ref{table1} shown in Fig. \ref{fig2}).  
 
\section{Properties of the debris disk candidates}\label{disks} 
 
We now discuss individual debris disk candidates in more detail.  
We arrive at a similar conclusion as \citet{Stauffer05}:
except for the 24 $\mu$m excess, the host stars for debris disks 
do not stand out as a group in any other properties.
We find however that excesses are more frequent among early-type
stars, and that the solar-type candidates are predominantly single stars.
 
\subsection{IRAC excesses}  
All disk candidates, except for HII 1095, 
HII 2425 and Pels 58, have been also imaged with IRAC. 
None was found to show excess at wavelengths $\leq8\mu$m, 
similar to cirrus-free non-excess members 
\citep[][2006 in preparation]{Stauffer05}. This property provides
yet another defense against false excesses from heated 
interstellar material, since in such cases we would expect
to detect aromatic emission in the 6 to 8$\mu$m range. 
 
\subsection{Model Spectral Energy Distributions}

\citet{Stauffer05} have modeled the spectral energy distribution (SED) from 0.3 to 30 $\mu$m
in HII 1101. They found that the 24 $\mu$m emission arises
from a 84 K disk with a central hole 13 $AU$ in radius.
Considering that $V-K$, IRAC and MIPS fluxes
of our strongest solar-type disk candidate HII 1797 are very similar to HII 1101,
it could be that a similar cold disk is observed in HII 1797 as well,
and probably much different from the ones around
the F9 dwarf in M 47 and M dwarf in NGC 2547 that have much stronger 24 $\mu$m excesses
\citep{Gorlova04,Young04}. Given the relatively few constraints,
a variety of alternative disk models are also likely to be compatible
with the SEDs of these stars.

\subsection{Binarity}\label{Binarity}  
 
One can expect two opposite effects of binarity on debris disks.
The circumstellar disks in binaries may be more truncated and
the inner part of the circumbinary disk better cleared
compared to the single stars, in which case one expects an anticorreletion
between excess and binarity. On the other hand, a companion can stir up
the debris disk prompting more collisions and greater dust production. 
In the field and for old ($>$ 1Gyr) stars we find that binary
systems may tend toward a higher incidence of 24um excess
(Trilling et al., in preparation). The Pleiades can in principle 
provide a useful comparison at a younger stellar age. 
We therefore discuss binarity for the Pleiades excess stars.

\textit{HII 2195} has been suspected to be a spectroscopic binary based on 
the scatter of the radial velocity measurements \citep{Liu91}. 
The number of observations however is too small to allow determination 
of the period. The speckle observations of \citet{Mason93} did not detect any companion 
within 0.035-1\arcsec\, and less than 3 mag fainter than the primary. 
\citet{Raboud98} consider this star to be single. 
The binarity status therefore remains to be confirmed. 
 
As with HII 2195, \textit{HII 1284} was identified as a spectroscopic binary 
without a period in \citet{Liu91}, but no companion was  
detected in the speckle observations of \citet{Mason93}.  
\citet{Raboud98} however consider it a single-lined spectroscopic 
binary and derive a mass for the companion of 0.86 M$_{\odot}$, corresponding to  
spectral type G-K. Also, the weak X-ray emission detected 
within 2\arcsec\, by \textit{Chandra} 
is consistent with emission from an ``inactive late-type companion''
\citep{Daniel02}. 
It is interesting that HII 1284 is also a photometrically variable star (V1210 Tau). 
If indeed it belongs to the $\gamma$ Dor pulsating type, 
with $\Delta V = \pm$ 0.02 mag and P $=$ 8$^{h}$ \citep{Martin00}, 
this status could explain some of the radial velocity variations 
currently assigned to the invisible spectroscopic companion.  
 
\textit{HII 1101} was suspected to be a photometric and spectroscopic binary 
in \citet{Soderblom93}, confirmed as a spectroscopic binary in \citet{Queloz98}. 
It was however considered a single star in \citet{Raboud98} 
and most recently in \citet{Stauffer05}. 
Clearly the above three stars deserve further spectroscopic  
monitoring to establish their binarity. 
 
\textit{HII 1380} was reported as a visual binary in \citet{Anderson66}, 
but binarity is not mentioned in more recent  
studies \citep{Liu91,Raboud98,Dommanget00,Mason93}. 
The position and photometric information  
in the Washington Double star catalog \citep{Mason01} indicate 
that the companion in question could be HII 1368. 
The proper motion of the latter, however, 
as well as its large separation (29\arcsec\,) 
make it an unlikely cluster member. 
 
The rest of the disk candidates -- \textit{HII 489, 514, 996, 1095, 1200, 1797, 2425} -- 
are apparently single stars, as indicated by radial velocity 
studies \citep{Liu91,Raboud98} and their absence in 
the Hipparcos Visual Double Stars Catalog \citep{Dommanget00}. 
\textit{Pels 58} is not mentioned in the above binarity studies, 
perhaps because it is located $1.3 \degr\,$ away from 
the core (the North-most square in Fig. \ref{fig1}). 
Nevertheless, the consistency of the different radial velocity estimates 
reported for it in \textsl{VizieR}, and the suggestion 
of \citet{Kharchenko04} to consider it a radial velocity standard 
indicate it to be a single star as well. 
 
We looked at the 2MASS images of the disk candidates 
for physical companions within 15\arcsec\, 
that could have been missed in the optical studies  
(15\arcsec\, at the Pleiades distance 
corresponds to the maximum separation set by 
the dynamical interactions with other cluster members). 
Also we looked for any other objects within the MIPS beam (6\arcsec\,) 
whose emission could be confused with the member flux. 
We found \textit{HII 2425} to have companion 11\arcsec\, north-east, which 
is visible in the 2MASS K band only.
\textit{HII 1101} has a 2MASS companion 9\arcsec\, south-east,  
too faint however to identify its nature from the NIR colors. 
None of the remaining disk candidates has 2MASS counterparts 
within 15\arcsec. 
 
Fig. \ref{fig5} shows that at 24 $\mu$m only \textit{HII 1284} reveals a companion  
(11\arcsec\, to the south-east, at \textit{RA=03:47:04.5}, \textit{DEC=+23:59:32.7}) with F$_{24}=2.1$ mJy. 
It is seen at the same location on the East and West sub-fields obtained few days apart. 
The absence of a 2MASS counterpart, meaning $K>14^{m}$, and the small reddening 
of HII 1284, indicate that, if a Pleiades member, this object would have to
be of very low luminosity (M6 or later) with a huge excess. 
The observed colors are  
too red for an optically thin debris disk, comparable only to excesses from primordial disks 
in much younger objects \citep[][Muzerolle et al. 2006 in preparation]{Lada06}. 
The companion to HII 1284 appears therefore to be an unrelated source. 
 
We conclude that the debris disk candidates 
in the Pleiades are predominantly single stars. 
A few cases -- HII 1284, HII 2195 and HII 1101 
are possible, but unconfirmed, spectroscopic binaries.
Thus our debris disk stars
should be directly comparable with the
field-star studies of \citet{Kim05} and \citet{Bryden06} that
apply to predominantly single-star samples. 
 
\subsection{Rotation} 
The distribution of Pleiades rotational velocities with temperature/mass 
is bimodal. The first group includes all spectral types and has the upper 
envelope of the $vsini$ distribution exponentially decaying toward later types, 
starting from above 300 km/s in B stars (e.g., Pleione), 
to 7 km/s in K, to 3 km/s in early Ms. The second 
group consists of late-type ultra-fast rotators,  
with velocities of 25-140 km/s \citep{Stauffer84,Soderblom93,Queloz98}.
All excess stars fall into the first rotation group
formed by the majority of members. 
With the limited number statistics per spectral bin,
we do not find significant correlation of 
the strength of the 24 $\mu$m excess with rotation. 
For example, the early members HII 2425 and HII 1380  
share similar spectral types and fast rotation, 
but their excesses are markedly different.
Half of the late-type ultra-fast rotators
fall in the region of heavy cirrus ($RA < 03^{h} 48^{m}$, Fig. \ref{fig1}),
making reliable measurement of their excesses impossible. The
remaining are HII 2034, HII 1516, HCG 154, and possibly
HII 174, none of which has an excess.
  
\subsection{X-rays} 
Recent X-ray studies by \textit{Chandra} \citep{Daniel02} 
and \textit{XMM-Newton} \citep{Briggs03} surveyed areas 
in the cluster core just eastward of the heavy cirrus. 
The \textit{XMM-Newton} field is a $30\arcmin\, \times30\arcmin\,$ 
area north-east of Alcyone and Maia. 
Except for the late K dwarfs HII 1110 and HII 1280, 
within this field we detect at 24 $\mu$m all 9 B9.5-K members  
detected also in X-rays, as well as two X-ray 
non-detections, the early A stars HII 1431 and HII 1028. 
None of these stars shows convincing IR excess. 
Two of them are possibly contaminated:
HII 1234 is embedded in a smooth 0.5\arcmin\, halo,  
while HII 1028 is somewhat extended for a point source. 
The \textit{Chandra} field is an  
adjacent $16\arcmin\, \times16\arcmin\,$ 
area between Alcyone and Merope. 
There are 8 B6-K6 common detections 
between MIPS and \textit{Chandra}, 
and two non-detections -- the A stars HII 1362 and HII 1375. 
Because this area is strongly affected by cirrus, 
only 3 objects have reliable 24 $\mu$m photometry. 
Two of them, the mid-F spectroscopic binaries HII 1122 and HII 1338, 
do not have 24 $\mu$m excess. 
The third one, HII 1284, A9, 
is one of the 12 debris disk candidates. 
The X-ray emission is presumably coming from the spectroscopic 
``inactive late-type'' companion \citep{Daniel02}. 
Among 5 objects with questionable photometry, 
3 may have 24 $\mu$m excesses: 
the Be star HII 980, A7 HII 956 and K5 HII 1061, 
all of which are suspected binaries. 
\textit{ROSAT} observations are less sensitive 
but more complete in terms of the covered area. 
\citet{Micela99} and \citet{Stauffer05} provide 
X-ray fluxes for all the solar-type debris candidates (with
upper limits for HII 1095 and HII 1200), 
and upper limits for B-A candidates (except for Pels 58). 

\begin{figure} 
\plotone{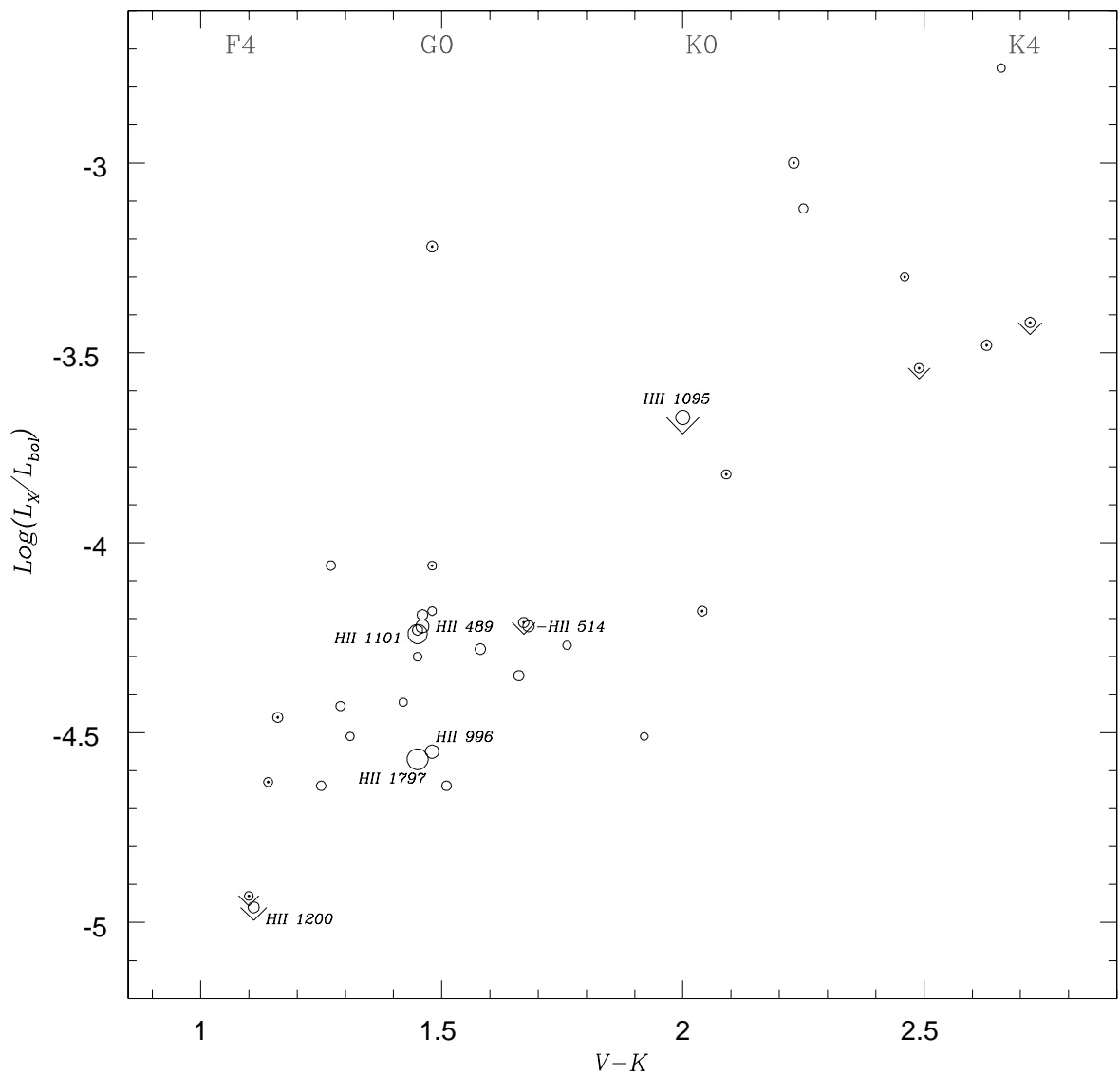} 
\caption{Solar-type members from Table \ref{table2} 
with X-ray luminosity fractions from \citet{Micela99} and \citet{Stauffer05}. 
\textit{Dotted circles:} binaries; \textit{downward arrows:} upper limits. 
Bigger circles correspond to bigger $K-[24]$ excess (defined as the deviation 
from the fit to the non-excess locus in \S\ref{excess}). 
Marked are disk candidates. 
The three strongest single X-ray emitters are 
the fast rotators HII 727, HII 174, \& HII 2034. 
}\label{fig6}   
\end{figure}

\citet{Chen05a} noticed a possible anti-correlation between X-ray activity 
and debris disk presence, interpreting 
it as evidence for dust clearing by stellar wind drag. 
To test this correlation, we plot X-ray luminosities 
for the solar-like Pleiades members from Table \ref{table2} 
in Fig. \ref{fig6}. 
The sizes of the circles represent the magnitude of the 24 $\mu$m excesses. 
As one can see, the disk candidates are not 
distinguishable from the diskless ones in this diagram
(in agreement with \citet{Stauffer05}). 
Although we do not confirm the \citet{Chen05a} finding,
a larger sample is needed for a better test. 
In addition, the Pleiades stars are significantly older
than the members of the Sco-Cen association studied by Chen et al.,
and the wind drag must be less effective because the stellar wind is weaker.

\section{Lifetime of the 24 $\mu$m excess}\label{lifet} 
 
We will now compare the excess rate for the Pleiades stars with results
from other studies. Various authors use slightly different criteria to define 
a significant excess, dictated mainly by the precision of their IR measurements, 
and split their samples into slightly different spectral type/color bins, 
which are usually limited by the volume and detection limits 
of a given survey. Spectral type and binarity information are not 
available for the majority of cluster stars either. 
For a preliminary comparison  
with other studies we define those Pleiades members to have excesses  
that deviate by more than 0.15 mag ($\sim$3$\sigma$) 
from the $K-[24]$ locus of non-excess members as derived 
in \S\ref{excess}. We calculate excess frequencies in the following spectral-type groups
defined with colors (dereddened where possible): 
1) early-type stars -- B--A, or $V-K$ $\leq$ 0.8; 
2) solar-type stars -- F--K3, 0.8 $<$ $V-K$  $\leq$ 2.7. 
We further compute excess fractions separately for single stars. 
In all cases only stars from Table \ref{table2} are used. 
For our total sample of non-emission \textit{early-type} Pleiades stars  
we obtain an excess fraction of 25\% (5 stars out of 20); 
for a subsample of single stars it is 33\% (3/9). 
For \textit{solar-type} stars we obtain correspondingly: 
10\% (4/40) based on our measurements only, 9\% (5/53) when we include the  
13 stars from \citet{Stauffer05}, 
and for the single stars only in the combined sample -- 17\% (5/30). 
Note that a 0.06 mag difference in [24] 
between us and \citet{Stauffer05} puts HII 514 just below the  
boundary between excess and non-excess objects. 
Our 0.15 mag excess threshold also eliminates 
HII 1200 as an excess candidate,
so we add only HII 1101 from \citet{Stauffer05} as having a confirmed excess.
 
In Table \ref{table3} we compile \textit{Spitzer} results 
on the 24 $\mu$m excess fractions in several young clusters 
and associations and in the field, separately for 
early-type and solar-type stars. 
They are binned into several age categories that roughly correspond 
to the main episodes of dust production in the Solar System -- 
clearing of the primordial disk and vigorous collisions between 
growing planetesimals within the first 10 Myr; 
rare collisions between terrestrial planet embryos  
(e.g., the Earth-Moon creation at 30 Myr); 
possible migration of giant planets 
that stir smaller bodies and initiate 
collisional cascades (Late Heavy Bombardment at 700 Myr); 
and finally a constant process of dust production 
from colliding asteroids and evaporation of comets. 
 
\begin{figure} 
\plotone{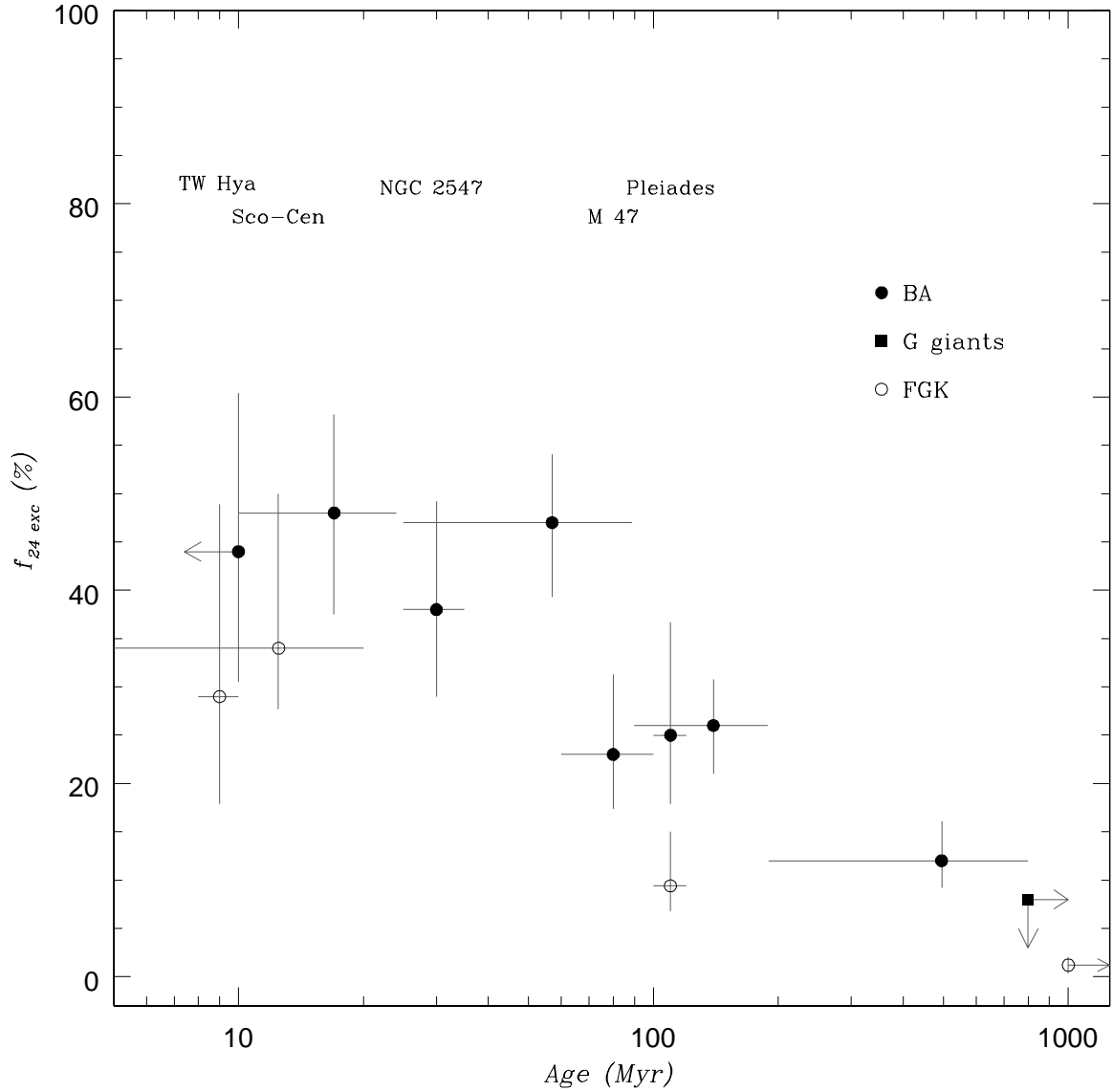} 
\caption{Decay of 24 $\mu$m excess fraction as a function of age, 
based on \textit{Spitzer} studies of stars in the young clusters 
and the field (Table \ref{table3}). \textit{Solid circles:} early-type stars; \textit{open circles:} solar-type 
stars; \textit{solid square:} G giants (descendants of A stars). 
}\label{fig7}   
\end{figure}

We derive another limit on debris disk excesses using G giants.
G giants are tracers of the post-main-sequence evolution of A stars, 
unlike cooler K-M giants which are a mixture  
of stars with a wide range of masses 
and at various late evolutionary stages. 
\citet{Jura90} searched the \textit{IRAS} Point Source Catalog 
for 60 $\mu$m excess emission in G giants. He found no 
confident excess among 36 giants. 
This allows estimation of an upper limit $\epsilon_{up}$  
on the true excess fraction, using the binomial distribution 
as appropriate for the small number statistics here. 
At a confidence level of 0.954 (that corresponds to $\pm 2 \sigma$ 
in the Gaussian approximation) 
$\epsilon_{up}$ is calculated from 
$\int_{0}^{\epsilon_{up}} 37(1-\epsilon)^{36}d\epsilon = 0.954$, 
resulting in $\epsilon_{up}=0.08$. 
Based on a larger sample (that includes fainter stars) \citet{Plets97} 
found the 60 $\mu$m excess fraction to be 0-3\% for G giants, and even less at 25 $\mu$m.
\citet{Kim01} reexamined a dozen $IRAS$ excesses around G-K giants with $ISO$
and found dust to extend at least a few thousand $AU$ (with T $<$ 100 K), which excludes
a debris disk interpretation. We therefore adopt 8\%
as a conservative limit on the 25 $\mu$m excess fraction
from debris disks around evolved stars of a few solar masses.
Data from Table \ref{table3} are plotted in Fig. \ref{fig7}. 

An important new result is the high level of excesses at 24 $\mu$m for 
the roughly solar mass stars in the Pleiades compared to older field dwarfs.
For the field dwarf sample, we have combined the work 
of \citet{Kim05}, \citet{Bryden06}, and Beichman et al. (2006, 
in preparation) into a total sample of 167 stars of solar mass and 
typical age $>$1 Gyr. Of these stars, two have excesses at 24 $\mu$m, 
compared with five of 53 solar mass stars in the Pleiades. We have used 
the binomial theorem to test the significance of this difference and 
find a probability of less than 0.3\%  that the two samples are drawn 
from the same distribution. That is, 24 $\mu$m excesses are more common 
at 100 Myr than at $>$1 Gyr at a statistically significant level.

The incidence of Pleiades 24 $\mu$m excesses appears to be higher still in stars
of $\sim$ 2.5 M$_\odot$ (A-stars) -- 25\% vs. 9\% in solar mass stars
(with a 96\% confidence).
Does this mean that debris disk masses are systematically greater in high-mass stars?
For example, a broad range of masses was found for primordial disks around low-mass stars
in \citet{Andrews05}, which may translate into varying planetesimal masses
and give rise to a great variation in debris dust production.
However, in the absence of photometry at longer wavelengths,
one can not rule out that the observed difference in excesses 
is due to the difference in the stellar luminosity or in the spectral energy distribution.
More luminous A stars are capable of heating up material 
to larger distances, resulting in a geometrically larger area
that emits at 24 $\mu$m. On the other hand, radiation pressure
which is the dominant dust removal mechanism around luminous stars,
is also a function of wavelength. Stellar mass and wind
are additional factors that enter the expressions for the forces
acting on the dust particle \citep[e.g.,][]{Chen05b}.
At this time it is not clear which of the these
effects is the most important cause of the difference.
While this observation awaits confirmation from other clusters,
the decline in disk fraction with time for both A- and solar-type stars
is similar. We thus qualitatively conclude that the disk evolution is similar 
over the 0.8 to 2.5 M$_{\odot}$ stellar mass range,
and that a significant number of stars within this range are still
producing debris at high levels in their planetary zones ($\sim$1 -- 20 $AU$) at 100 Myr of age. 

The apparent difference between field A-stars and field solar-type stars arises
to first order simply because the A-stars are systematically younger.
We can therefore use the solar-mass stars as an indication of the 
evolution of debris disks past the main sequence lifetime of A stars. 
Twenty two of the sample of 167 solar-mass stars discussed above are 
detected at 70 $\mu$m (in general, the observations are deep enough to 
detect the stellar photospheres, although often at only low signal to 
noise). The rate of detection at this wavelength is 13$\pm$3\%, 
compared with a rate of $\sim$1\% at 24 $\mu$m.
Su et al. 2006, to be submitted to ApJ, demonstrates for the field A stars of various ages
and \citet{Smith06} for nearby B-K young stars,
that the 70 $\mu$m excess fraction decays slower than the 24 $\mu$m one.
We conclude that the  decline of the 24 $\mu$m emission with age corresponds to the decline of 
the frequency of collisions or disruptive events within the inner parts 
of the planetary systems, corresponding to the asteroidal zone of the 
solar system. The outer parts detected at 70 $\mu$m, corresponding to the 
Kuiper Belt zone, evolve more slowly.

Terrestrial planets had formed in the Solar system at $\sim$30 Myr \citep{Kleine02}
and presumably largely cleared material in the inner few $AU$ region
that we are probing at 24 $\mu$m. 
However, in different planetary system architectures, Moon-sized bodies within 1 $AU$
may still survive by the Pleiades age of $\sim$115 Myr \citep{Basri96}.
Occasional collisions could give rise
to intermittent excesses at 24 $\mu$m \citep{Kenyon05}.
Another source of warm dust at this age could be 
occasional disturbances of the asteroid or cometary belts
by migrating giant planets. 
Dynamical simulations show that giant planets can experience
strong interactions near the mean motion resonances,
resulting in dramatic episodes of scattering and producing a 
collisional cascade of the smaller bodies,
perhaps explaining the Late Heavy Bombardment
at 700 Myr in the inner Solar system \citep{Gomes05,Strom05}.
 
\citet{Kenyon05} have quantified this picture in a series of simulations 
for a solar-type and an A-type star. The predicted evolution of the 24 $\mu$m 
excess can be directly compared to observations from our Figs \ref{fig4} and \ref{fig7}. 
After 10 Myr an overall decline in 24 $\mu$m excess is predicted, 
but with levels that are always bigger for the more luminous A stars. 
Indeed, the \textit{average} excess is observed to decline with age 
for both early and late-type stars, and the fraction of 
detected excesses appears smaller in the solar-type stars 
compared to the early-type ones (Fig. \ref{fig7}). 
But how can we explain the observed \textit{range} of excesses 
between stars of similar age and mass? 
For solar-type stars, the simulations show it can be explained by    
collisions within a few $AU$ that 
produce short-lived excesses in amounts much bigger than the weak preexisting level. 
For A stars, however, the simulations behave differently:
the 24 $\mu$m emission comes from 
distances up to 100 $AU$, and the excess from a single collision 
never competes with the excess from the bulk of the disk 
(compare Figs. 4 and 8 in \citet{Kenyon05}).
This result also appears to contradict the observed extent of the
Vega system \citep{Su05}. Therefore, explaining disk behavior as
a function of stellar mass is an area requiring further investigation. 

\section{Conclusions}

We have conducted a photometric survey for warm dusty disks
in the Pleiades, the nearest medium-age ($\sim$100 Myr) open cluster.
At this age, terrestrial planet formation should be complete, 
and the inner few $AU$ region is expected to be largely cleared of 
dust.
The dust however can be temporarily replenished,
either from occasional asteroid collisions
or from the outer regions where giant planet migration may still be occurring.
Indeed, we find 24 $\mu$m excess emission around nine B9-K0
non-emission stars. The fraction of B-A stars with excesses is 25\%,
comparing well to a similar age cluster M 47 \citep{Gorlova04}.

Because of the proximity of the Pleiades, for the first time we
can explore excesses in solar-mass stars down to the photospheric limit.
Combined with the \citet{Stauffer05} sample,
we find that the incidence of 24 $\mu$m excesses for these stars, at
10\%, is significantly higher than that for old field solar-mass stars, 
1\%. Thus, the clearing of debris in the planetary zone ($\sim$1 -- 20 $AU$)
for these stars takes a similar time as the similar process
in A-stars, of order 100 million years. 
Comparing with the incidence of 70 $\mu$m excess in $>$1 Gyr old
solar mass stars, 13$\pm$3\%, it appears that the outer, Kuiper Belt - like 
zones of planetary debris clear much more slowly that the planetary zones seen at 24 $\mu$m.
 
\acknowledgments 

We would like to thank the referee for comments that helped clarify the paper.
This work was supported by JPL/Caltech under contract 1255094.

\clearpage   
\begin{deluxetable}{lrccrrcl} 
\tablecolumns{10}  
\tablewidth{0pt} 
\tabletypesize{\tiny} 
\tablecaption{Pleiades members with contaminated 24 $\mu$m photometry\label{table1}} 
\tablehead{ \colhead{Name} & \colhead{$(V-K)_{0}$\tablenotemark{1}} & \colhead{[24] flag\tablenotemark{2}} & 
\colhead{SpT} & \colhead{$A_{V}$} & \colhead{vsini} & 
\colhead{Binary?} & \colhead{2MASS}\\ 
\colhead{} & \colhead{mag} & \colhead{ } & \colhead{} & \colhead{mag} &  
\colhead{km/s} & \colhead{} & \colhead{} 
} 
\startdata 
\textbf{HII 0468} & \textbf{-0.25} & \textbf{h}                     &\textbf{B6IIIe} & \textbf{0.04} & \textbf{    220}&     \textbf{SB1?} &  \textbf{03445253+2406478} \\  
 HII 0785 &   -0.24 &  h                     & B7III &    0.16 &      40 &          OccB        &     03454960+2422037 \\  
 \textbf{HII 0541} &\textbf{   -0.24} &\textbf{h}                     & \textbf{B8V} & \textbf{0.09} & \textbf{245} &   \textbf{SB?} & \textbf{03450973+2450213} \\  
 \textbf{HII 0980} &  \textbf{-0.22} &\textbf{me}                   &\textbf{B6IVe} &  \textbf{0.21} &   \textbf{275} & \textbf{B?} &   \textbf{03461958+2356541} \\  
 HII 0447 &   -0.20 &  h                     &  B7IV &    0.19 &     260 &         OccB         &     03444821+2417222 \\  
\textbf{HII 0817} & \textbf{-0.16} & \textbf{h}                    & \textbf{B8V} &  \textbf{0.16} & \textbf{220} &                  &  \textbf{03455447+2433162 }\\  
 \textbf{HII 0859} & \textbf{-0.13} & \textbf{h}                    & \textbf{B9V} & \textbf{0.15} & \textbf{250} &                  &  \textbf{03460288+2431403} \\  
 \textbf{HII 1234} & \textbf{-0.10} & \textbf{h}                      &\textbf{B9.5V} &  \textbf{0.26} &  \textbf{260} &    VB              &  \textbf{03465940+2431124 }\\  
 \textbf{HII 1375} &  \textbf{0.03} &\textbf{me}                    & \textbf{A0V} &\textbf{0.00} &   \textbf{160} & \textbf{SB} &   \textbf{03472103+2406586} \\  
HII 1397 &    0.12 & cc                     &   A2V &    0.00 &    $<$10&          SB1, VB 6.26\arcsec &     03472440+2354529 \\  
\textbf{HII 1028} &  \textbf{0.18} & \textbf{me}                   & \textbf{A2V} & \textbf{0.05} &  \textbf{110} &   \textbf{SB?} &   \textbf{03462728+2415181} \\  
 \textbf{HII 0652} & \textbf{0.35} & \textbf{h}                     & \textbf{A3V} & \textbf{0.26} &  \textbf{235} &  \textbf{SB?}      &  \textbf{03452613+2402065} \\  
 \textbf{HII 0956} &   \textbf{0.76} & \textbf{me}                  &   \textbf{A7V} & \textbf{0.10} &  \textbf{150} &   \textbf{SB?, VB} &  \textbf{03461600+2411234}\\  
 HII 0745 &    1.05 & b                      &   F5V &    0.40 &      65 &   SB1?            &     03454138+2417189 \\  
 HII 0476 &    1.19 &  b                     &    F9 &    0.87 &      21 &            PHB, SB &     03445383+2355165 \\  
 HII 0761 &    1.35 & b                      &    G2 &    0.50 &      11 &              SB1&     03454440+2413132 \\  
 HII 1215 &    1.46 &  b                     &    G0 &    0.11 &       6 &                  &     03465373+2335009 \\  
 \textbf{HII 0320} &\textbf{1.63} &\textbf{cc}                    &\textbf{G5} &\textbf{0.61} &\textbf{11} & \textbf{SB1} &\textbf{03442050+2446222} \\  
\textbf{HII 0659} & \textbf{1.68} &\textbf{b}                    &  \textbf{G4} &  \textbf{0.74} &   \textbf{12} &                  &  \textbf{03452597+2325487 }\\  
 HII 0746 &    1.72 & b                      &    G5 &    0.20 &       5 &              SB?&     03454184+2425534 \\  
  HII 0870 &    1.78 &  b                     &  G4.5 &    1.77 &      10 &          VB 0.51\arcsec &     03460275+2344146 \\  
 HII 1275 &    1.78 & b                      &    G8 &    0.15 &       6 &                  &     03470141+2329419 \\  
 HII 1032 &    1.84 & b                      &    G8 &    0.11 &      36 &                  &     03462841+2426021 \\  
 \textbf{HII 0430} & \textbf{1.84} & \textbf{b}                     &\textbf{G8} & \textbf{0.09} & \textbf{7} &                  & \textbf{03444398+2413523} \\  
 HII 1136 &    2.23 & b                      &    G8 &    0.68 &      68 &                  &     03464024+2329520 \\  
 HII 1124 &    2.24 & b                      &    K1 &    0.25 &       6 &                  &     03463938+2401468 \\  
 HII 0522 &    2.24 & b                      &    K2 &    0.00 &       4 &              SB1 &     03450326+2350219 \\  
 HII 0625 &    2.27 & b                      &    K0 &    1.15 &      94 &                  &     03452118+2343389 \\  
 \textbf{HII 0738} & \textbf{2.30} &\textbf{b}                      &  \textbf{G9} & \textbf{1.12} &  \textbf{50} & \textbf{VB 0.50\arcsec} & \textbf{03453940+2345154 }\\  
 HII 1039 &    2.40 & b                      &    K1 &    0.84 &    $<$5 &                  &     03462777+2335337 \\  
HII 1332 &    2.52 & b                      &   K4  &    0.00 &     5   &                  &     03471352+2342515 \\     
 HII 0636 &    2.55 & b                      &       &          &       4 &                  &     03452219+2328182 \\  
 HII 0883 &    2.80 & b                      &       &          &       4 &                  &     03460689+2433461 \\  
 HII 0882 &    2.98 &  b                     &    K3 &    0.15 &      65 &                  &     03460412+2324199 \\  
 HII 1348 &    2.98 & b                      &   K5 &    0.00 &       5 &       SB2  &     03471806+2423267 \\  
 HII 0451 &    3.09 &  b                     &    K5 &    0.06 &       6 &                  &     03445017+2454400 \\  
 HII 0559 &    3.11 &  b                     &       &          &      65 &                  &     03451352+2505159 \\  
 \textbf{HII 1061} & \textbf{3.20} & \textbf{b}                    & \textbf{K5} &  \textbf{0.62} &  \textbf{7} &  \textbf{VB 0.32\arcsec} &  \textbf{03463117+2407025 }\\  
 HII 0686 &    3.23 & b                      &   K7  &    0.06 &     64  &                  &     03453293+2418116 \\ 
HII 1531 &    3.27 &  b                     &   K7.5&    0.00 &     50  &                  &     03474143+2358190 \\ 
 HII 1081 &    3.44 & b                      &   K6  &    0.40 &    $<$10&                  &     03463287+2318191 \\   
 HII 1355 &    3.45 & b                      &    K6n &    0.34 &      12 &          VB 1.26\arcsec &     03471814+2402114 \\  
 HII 1103 &    3.73 & b                      &   K7  &    0.34 &     18  &                  &     03463532+2324424 \\  
HCG 0196 &    3.82 &  b                     &       &          &         &                  &     03453903+2513278 \\  
HCG 0152 &    4.03 &  b                     &       &          &         &                  &     03443006+2535470 \\  
HCG 0277 &    4.84 &  b                     &       &          &         &                  &     03473345+2341330 \\  
\enddata 
\tablenotetext{1}{Dereddened where SpT avaialble} 
\tablenotetext{2}{me -- \textbf{m}arginally \textbf{e}xtended source, FWHM $=$ 6.5 -- 7\arcsec; h -- a \textbf{h}alo-like 
extended smooth cirrus, with or without point source on top; b -- within strongly non-uniform \textbf{b}ackground; 
cc -- \textbf{c}lose \textbf{c}ompanion } 
\tablecomments{Objects sorted according $(V-K)_{0}$. In bold are sources with nominal 24 $\mu$m excesses.} 
\end{deluxetable}

\clearpage   
\begin{deluxetable}{lrrrrcrrcl} 
\tablecolumns{10}  
\tablewidth{0pt} 
\tabletypesize{\tiny} 
\tablecaption{Pleiades members with reliable 24 $\mu$m photometry\label{table2}} 
\tablehead{ \colhead{Name} & \colhead{$(V-K)_{0}$\tablenotemark{1}} & \colhead{$(K-[24])_{0}$\tablenotemark{1}} & 
\colhead{$F_{24}$} & \colhead{$\sigma F_{24}$} & \colhead{SpT} & \colhead{$A_{V}$} & \colhead{vsini\tablenotemark{2}} & 
\colhead{Binary?} & \colhead{2MASS}\\ 
\colhead{} & \colhead{mag} & \colhead{mag} & \colhead{mJy} & \colhead{mJy} & \colhead{} & \colhead{mag} &  
\colhead{km/s} & \colhead{} & \colhead{} 
} 
\startdata 
 HII 2168 &   -0.32 &    0.02 &     210.15 &    3.10 & B8III &    0.09 &     215 &   VB, SB1? &     03490974+2403121  \\  
 HII 0563 &   -0.25 &   -0.18 &      98.66 &    0.91 &   B6V &    0.07 &     135 &   SB1? &     03451250+2428021  \\  
 HII 1823 &   -0.18 &   -0.16 &      38.06 &    0.39 &   B8V &    0.08 &     270 &        &     03482081+2325165  \\  
\textbf{HII 2425} &\textbf{-0.14} &\textbf{0.69} & \textbf{45.60} &\textbf{0.67} &\textbf{B9V} &\textbf{0.10} &\textbf{310} &        &\textbf{03494353+2342427} \\  
 HII 1431 &   -0.05 &    0.01 &      17.26 &    0.22 &   A0V &    0.29 &      40 &    SB2 &     03472945+2417180  \\  
 HII 0717 &    0.02 &   -0.15 &      15.17 &    0.24 &   A1V &    0.60 &      15 &   VB, SB? &     03453777+2420083  \\  
\textbf{HII 2181} &\textbf{0.04} &\textbf{2.30} &\textbf{652.41} &\textbf{3.00} &\textbf{B8Vpe} &\textbf{0.13} &\textbf{340:}&\textbf{VB} &\textbf{03491121+2408120} \\ 
 HII 1084 &    0.05 &   -0.06 &      11.53 &    0.13 &   A0V &    1.12 &     150 & SB2?     &     03463420+2337264  \\  
\textbf{HII 1380} &\textbf{0.10} &\textbf{0.17} &\textbf{14.85} &\textbf{0.27} &\textbf{A1V} &\textbf{0.00} &\textbf{235} & &\textbf{03472096+2348121} \\  
\textbf{HII 1432} &\textbf{0.18} &\textbf{0.85} &\textbf{1414.24} &\textbf{20.84} &\textbf{B7IIIe} &\textbf{0.06} &\textbf{220} &  \textbf{SB1}   &\textbf{03472908+2406184} \\  
 HII 0804 &    0.18 &   -0.06 &       8.40 &    0.14 &   A2V &    0.38 &     170 &  SB1      &     03455163+2402200  \\  
\textbf{Pels 58} &\textbf{0.27} &\textbf{1.07} &\textbf{19.24} &\textbf{0.34} & \textbf{A3} & \textbf{0.21} &\textbf{95} &        &\textbf{03455913+2523549} \\  
 HII 1876 &    0.39 &   -0.00 &      17.26 &    0.19 &   A1V &    0.00 &     105 &   PHB, SB? &     03483009+2420441  \\  
 HII 1362 &    0.43 &    0.02 &       6.71 &    0.11 &    A7 &    0.21 &     $<$12 &        &     03471935+2408208  \\  
\textbf{HII 2195} &\textbf{0.49} &\textbf{0.25} &\textbf{8.59} &\textbf{0.09} &\textbf{A7V} &\textbf{0.06} &\textbf{160} &\textbf{SB?} &\textbf{03491219+2353126} \\  
 HII 1384 &    0.50 &   -0.04 &      11.32 &    0.17 &   A4V &    0.18 &     215 &        &     03472405+2435184  \\  
 HII 0531 &    0.58 &    0.03 &       6.20 &    0.12 &   Am? &    0.30 &      75 &        &     03450653+2415486  \\  
\textbf{HII 1284} &\textbf{0.65} &\textbf{0.30} &\textbf{8.56} &\textbf{0.13} & \textbf{A9V} & \textbf{0.09} & \textbf{100} &\textbf{SB1?}&\textbf{03470421+2359426}\\  
 HII 1266 &    0.67 &   -0.05 &       8.40 &    0.20 &   A9V &    0.31 &      95 &     VB &     03470354+2449117  \\  
 HII 0344 &    0.68 &    0.03 &       7.71 &    0.16 &   A8V &    0.02 &     200 &        &     03442570+2423408  \\  
 HII 0697 &    0.74 &    0.04 &       6.60 &    0.09 &    A9 &    0.21 &      75 &        &     03453445+2427478  \\  
 HII 1762 &    0.80 &   -0.05 &       8.10 &    0.10 &   A9V &    0.13 &     180 & SB2    &     03481354+2419063  \\  
 HII 0975 &    $<$0.42: &   $<$-0.05: &       3.36 &    0.08 &   F &    $>$1.86: &      32 &    PHB &     03461799+2329119  \\  
 HII 0530 &    0.95 &    0.08 &       4.95 &    0.10 &    F3 &    0.00 &     $<$12 &        &     03450528+2342097  \\  
 HII 1122 &    0.97 &    0.09 &       4.27 &    0.09 &    F4 &    0.15 &      29 &   SB2 &     03463932+2406116  \\  
 HII 0605 &    0.99 &   -0.05 &       5.20 &    0.09 &    F3 &    0.18 &      80 &    SB1 &     03452085+2455194  \\  
 HII 2345 &    0.99 &   -0.07 &       4.21 &    0.17 &    F4 &    0.08 &     130 &        &     03493272+2322494 \\
 HII 1309 &    1.06 &   -0.03 &       3.49 &    0.11 &    F6 &    0.14 &      85 &        &     03471005+2416360  \\  
 HII 1338 &    1.10 &   -0.04 &       7.21 &    0.10 &    F3 &    0.11 &      10:&   SB2 &     03471656+2407420  \\  
 \textbf{HII 1200} & \textbf{1.11} &\textbf{0.10} & \textbf{3.11} &\textbf{0.05} &\textbf{F6} &\textbf{0.29} &\textbf{14}  &            &\textbf{03465053+2314211\tablenotemark{3,4}}  \\  
 HII 1912 &    1.14 &   -0.03 &       5.63 &    0.12 &    F4 &    0.16 &      75 &     VB &     03483480+2410523  \\  
 HII 1726 &    1.16 &    0.04 &       5.16 &    0.10 &    F7 &    0.17 &      13 &     IRB 0.57\arcsec &  03480718+2408315  \\  
 HII 1613 &    1.25 &    0.02 &       2.78 &    0.08 &    F8 &    0.07 &      20 &        &     03475252+2356286  \\  
 HII 0727 &    1.27 &    0.01 &       3.61 &    0.08 &    F9 &    0.16 &      50 &        &     03454016+2437380  \\  
 HII 0405 &    1.29 &    0.00 &       2.89 &    0.08 &    F9 &    0.03 &      18 &        &     03444075+2449067  \\  
 HII 1856 &    1.31 &   -0.06 &       2.39 &    0.07 &    F8 &    0.05 &      15 &        &     03482616+2402544  \\  
 HII 2506 &    1.37 &   -0.06 &       2.08 & 0.05 &    F8 &    0.10 &      14  &            &     03495648+2313071\tablenotemark{3}  \\  
 AK 1B 146 &    1.40 &   -0.07 &       3.56 & 0.06 &    F8 &    0.02 &      12/9    & SB2    &     03435067+2516081 \\
 HII 3179 &    1.42 &   -0.05 &       2.43 & 0.05 &    G0 &    0.00 &       5  &            &     03515685+2354070\tablenotemark{3}  \\  
\textbf{HII 1797} &\textbf{1.45} &\textbf{0.47} &\textbf{3.61} &\textbf{0.08} &\textbf{F9} & \textbf{0.03} &\textbf{20} &        &\textbf{03481691+2338125}  \\  
\textbf{HII 1101} &\textbf{1.45} &\textbf{0.42} &\textbf{3.32} &\textbf{0.05} &\textbf{F9.5} &\textbf{0.05} &\textbf{19} & &\textbf{03463878+2457346\tablenotemark{3}} \\  
 HII 1207 &    1.45 &    0.01 &       1.90 &    0.08 &  G0:V &    0.09 &       5 &        &     03465491+2447468  \\  
 HII 1924 &    1.45 &    0.06 &       2.20 &    0.07 &    G0 &    0.01 &      14 &        &     03483451+2326053  \\  
 HII 2786 &    1.45 &   -0.03 &       2.02 & 0.05 &       &         &      22  &            &     03504007+2355590\tablenotemark{3}  \\  
 HII 0923 &    1.46 &    0.08 &       2.72 &    0.08 &    G0 &    0.05 &      18 &        &     03461005+2320240  \\  
\textbf{HII 0489} & \textbf{1.46} &\textbf{0.21} & \textbf{2.55} &\textbf{0.09} &\textbf{G0} & \textbf{0.11} &\textbf{18} &        &\textbf{03445639+2425574} \\  
 HII 0120 &    1.48 &    0.01 &       1.71 & 0.05 &    G1 &    0.26 &       9  &       SB1? &     03433195+2340266\tablenotemark{3}  \\  
 HII 0152 &    1.48 &    0.13 &       1.84 & 0.05 &    G1 &    0.14 &      11  &            &     03433772+2332096\tablenotemark{3,5} \\  
% HII 0250 &    1.48 &    0.10 &       1.91 & 0.05 &    G1 &    0.18 &       6  &            &     03440424+2459233\tablenotemark{3} \\  
 HII 0250 &    1.48 &   -0.02 &       1.73 &    0.06 &    G1 &    0.18 &       7 &    SB? &            03440424+2459233  \\  
 HII 0293 &    1.48 &   -0.01 &       1.77 &    0.05 &    G1 &    0.28 &       7 &        &            03441391+2446457  \\  
% HII 0314 &    1.48 &   -0.02 &       2.00 & 0.05 &    G1 &    0.26 &      42  &          &     03442008+2447461\tablenotemark{3}  \\  
 HII 0314 &    1.48 &    0.10 &       2.26 &    0.09 &    G1 &    0.20 &      38 &  SB1?  &           03442008+2447461  \\  
\textbf{HII 0996} & \textbf{1.48} & \textbf{0.22} & \textbf{2.41} & \textbf{0.07} &\textbf{G1} &\textbf{0.02} &\textbf{12} &        &\textbf{03462267+2434126}  \\  
 HII 1015 &    1.48 &   -0.03 &       1.79 & 0.05 &    G1 &    0.08 &      10  &            &     03462735+2508080\tablenotemark{3}  \\  
 HII 1794 &    1.51 &    0.03 &       2.09 &    0.09 &      &     &      11 &            &     03481712+2353253  \\  
 HII 1182 &    1.53 &    0.01 &       1.95 & 0.05 &    G5 &    0.00 &      16  &   IRB 1.14\arcsec &     03464706+2254525\tablenotemark{3}  \\  
 HII 0739 &    1.58 &    0.02 &       5.00 &    0.08 &    G0 &    0.06 &      14 &    PHB &     03454211+2454215  \\  
 HII 1514 &    1.58 &    0.09 &       2.07 &    0.07 &    G5 &    0.00 &      14 &        &     03474044+2421525  \\  
 HII 2341 &    1.66 &    0.07 &       1.61 &    0.07 &    G4 &    0.06 &       3 &        &     03493312+2347435  \\  
 HII 1117 &    1.67 &    0.07 &       3.03 &    0.09 &    G6 &    0.00 &     6/4 &   SB2 &     03463767+2347159  \\  
\textbf{HII 0514} &\textbf{1.68} &\textbf{0.12} & \textbf{1.97} &\textbf{0.05} &       &         & \textbf{10}  &            & \textbf{03450400+2515282\tablenotemark{4}} \\  
 HII 2644 &    1.76 &   -0.03 &       1.32 & 0.05 &       &         &       4  &            &     03502089+2428003\tablenotemark{3} \\  
 HII 3097 &    1.80 &    0.05 &       1.66 & 0.05 &       &         &      15  &        SB1 &     03514044+2458594\tablenotemark{3}  \\  
 HII 0571 &    1.85 &    0.08 &       1.64 &    0.05 &    G8 &    0.21 &       8 &    SB1 &     03451534+2517221  \\  
 HII 2311 &    1.92 &   -0.08 &       1.15 &    0.05 &      &     &       6 &            &     03492873+2342440  \\  
\textbf{HII 1095} &\textbf{2.00} &\textbf{0.27} &\textbf{1.29} &\textbf{0.08} &\textbf{K0} &\textbf{0.18} &\textbf{4} &        &\textbf{03463777+2444517}\\  
 HII 0173 &    2.04 &    0.06 &       2.28 &    0.05 &    K0 &    0.00 &     8/6 &    SB2 &            03434841+2511241  \\  
% HII 0173 &    2.06 &   -0.00 &       2.12 & 0.05 &   K0 &    0.00 &     8/6  &        SB2 &     03434841+2511241\tablenotemark{3}  \\  
 HII 2278 &    2.09 &    0.03 &       2.21 & 0.05 &       &         &       7  &   VB 0.37\arcsec &     03492570+2456154\tablenotemark{3}  \\  
 HII 2027 &    2.12 &    0.01 &       2.20 &    0.06 &    K0 &    0.00 &       6 &    SB2, IRB 0.1\arcsec&     03484894+2416027  \\  
 HII 0174 &    2.25 &    0.03 &       1.34 & 0.04 &       &         &      90: &            &     03434833+2500157   \\  
 HII 2147 &    2.23 &    0.12 &       2.94 &    0.07 &    G9 &    0.00 &    7/11 &    SB2 &     03490610+2346525  \\  
% HII 2147 &    2.26 &    0.01 &       2.64 & 0.05 &   G9 &    0.00 &    7/11  &        SB2 &     03490610+2346525\tablenotemark{3}  \\  
 HII 2881 &    2.46 &   -0.02 &       1.70 & 0.05 &    K2 &    0.03 &      10  &   IRB 0.08\arcsec &     03505432+2350056\tablenotemark{3}  \\  
 HII 1298 &    2.49 &    0.05 &       0.89 &    0.07 &      &     &       6 &    IRB 1.18\arcsec &     03470678+2342546  \\  
 HII 1100 &    2.63 &    0.12 &       1.45 &    0.08 &    K3 &    0.25 &       5 &  VB 0.78\arcsec &     03463726+2420366  \\  
 HII 2034 &    2.66 &   -0.01 &       0.73 &    0.06 &  K2.5 &    0.00 &      75 &        &     03484932+2358383  \\  
 HII 0885 &    2.72 &    0.11 &       1.45 &    0.05 &    K3 &    0.00 &       6 &   IRB 0.87\arcsec &     03460776+2452004  \\  
 HCG 0131 &    2.95 &   -0.02 &       0.70 &    0.05 &      &     &         &            &     03440282+2539228  \\  
 HCG 0132 &    2.99 &    0.11 &       0.93 &    0.05 &      &     &         &            &     03440448+2551226  \\  
 HII 1653 &    3.34 &    0.20 &       0.76 &    0.07 &    K6 &    0.00 &      21 &        &     03475973+2443528  \\  
 HCG 0312 &    3.40 &    0.42 &       0.58 &    0.05 &       &     &     6.5 &            &     03481729+2430160  \\  
 HII 1516 &    3.64 &    0.33 &       0.73 &    0.06 &       &     &     105 &            &     03474037+2418071  \\  
 HCG 0154 &    3.98 &    0.10 &       0.79 &    0.06 &       &      &   75    &  IRB 4.61\arcsec\tablenotemark{6} &     03442729+2450382  \\  
 HCG 0354 &    4.55 &    0.60 &       0.63 &    0.06 &       &      &         &            &     03490585+2344232  \\  
\enddata 
\tablenotetext{1}{Dereddened where spectral type available} 
\tablenotetext{2}{The two values are for two components in SB2} 
\tablenotetext{3}{24 $\mu$m data from \citet{Stauffer05} } 
\tablenotetext{4}{Excess object according to \citet{Stauffer05}, but the excess is below our adopted 0.15 mag threshold} 
\tablenotetext{5}{Big offset between 2MASS and MIPS24 positions, likely contaminated with a background object} 
\tablenotetext{6}{Secondary is a background star}
\tablecomments{Objects sorted according to $(V-K)_{0}$. In bold are 24 $\mu$m excess candidates.} 
\end{deluxetable}  
 
\clearpage   
 
\begin{deluxetable}{lrlrcrcl} 
\rotate 
\tablecolumns{8}  
\tablewidth{0pt} 
\tabletypesize{\tiny} 
\tablecaption{24 $\mu$m excess fraction as a function of age\label{table3}} 
\tablehead{ \colhead{Epoch in the Solar system} & \colhead{Age} & \colhead{Cluster} & \multicolumn{4}{c}{Excess fraction} & \colhead{Ref.}  \\ 
\colhead{} & \colhead{} & \colhead{} & \multicolumn{2}{c}{B--A} & \multicolumn{2}{c}{F--G--K} & \colhead{} 
} 
\startdata 
Transition from &$<$10 Myr    &           & 44\%  & 4/9    &                           &       & \citet{Rieke05} \\ 
accretion to &5--20 Myr    & Sco-Cen   &       &        & 34--50\%\tablenotemark{1} & 14/41 & \citet{Chen05a} \\ 
debris stage &8--10 Myr    & TW Hya    &    &     & 29\%                         & 2/7  &\citet{Low05} \\ 
\\ 
Terrestrial & 10--24 Myr   &           & 48\%  & 10/21  &                              &      & \citet{Rieke05} \\ 
planet & 25--35 Myr   & NGC 2547  & 38\%  & 8/21   &                          &  & \citet{Young04,Rieke05} \\ 
formation & 25--89 Myr   &           & 47\%  & 20/43  &                              &      & \citet{Rieke05} \\ 
\\ 
Giant planet         & 90--189  Myr &           & 26\%  & 28/108 &                              &      & \citet{Rieke05} \\ 
migration, & 60-100 Myr   & M 47      &  23\% & 8/35   &  &  & \citet{Gorlova04} \\ 
interaction    & 100--120 Myr & Pleiades  & 25\%  & 5/20   & 9\%                         & 5/53 & this work and \citet{Stauffer05} \\ 
\\ 
Late heavy bombardment,& 190-800 Myr   &           &  12\% & 10/85  &                              &      & \citet{Rieke05} \\ 
asteroid grinding,&  $>$1 Gyr    &           &       &        & 1.2\%                        & 2/167 & \citet{Beichman05b,Kim05}; \\ 
destruction of comets&   &           &       &        &                        &  & \citet{Bryden06}; Beichman et al. 2006 in prep.\\ 
\\ 
Giant stage,                              &$>$800 Myr & & $<$8\% & 0/36 & & & \citet{Jura90} (G giants) \\ 
evaporating of comets in the Kuiper belt,  & & & & & & & \\ 
condensation of stellar wind & & & & & & & \\ 
\enddata 
\tablenotetext{1}{The observed level of 34\% is due to contamination by interlopers, accounting for them can bring the excess 
fraction to 50\%} 
\end{deluxetable}  
 
\clearpage 
 
\begin{thebibliography}{} 
\bibitem[Anderson, Stoeckly, \& Kraft(1966)]{Anderson66} Anderson, C. M., Stoeckly, R., \& Kraft, R. P.  1966, \apj, 143, 299 
\bibitem[Andrews \& Williams(2005)]{Andrews05} Andrews, S. M. \& Williams, J. P.  2005, \apj, 631, 1134
\bibitem[Arny(1977)]{Arny77} Arny, T., 1977, \apj, 217, 83
\bibitem[Backman \& Paresce(1993)]{Backman93} Backman, D. E. \& Paresce, F.  1993, in Protostars 
 and Planets III, eds. E. Levy \& J. I. Lunine (Tucson: Univ. Arizona Press), 1253 
\bibitem[Backman, Dasgupta \& Stencel(1995)]{Backman95} Backman, D. E., Dasgupta, A., \& Stencel, R. E.  1995, 
 \apj, 450, L35 
\bibitem[Baraffe et al.(1998)]{Baraffe98} Baraffe, I., Chabrier, G., Allard, F., \& Hauschildt, P. H. 
  1998, \aap, 337, 403 
\bibitem[Basri, Marcy, \& Graham(1996)]{Basri96} Basri, G., Marcy, G. W., \& Graham, J. R. 1996, \apj, 458, 600 
\bibitem[Beichman et al.(2005)]{Beichman05b} Beichman, C. A. et al.  2005, \apj, 622, 1160 
\bibitem[Beichman et al.(2006)]{Beichman06} Beichman, C. A., Tanner, A., Bryden, G., Stapelfeldt, K. R.,
 Werner, M. W., Rieke, G. H., Trilling, D. E., Lawler, S., \& Gautier, T. N.  2006,  ApJ, in press (astro-ph/0601468)
\bibitem[Bessell \& Brett(1988)]{Bessell88} Bessell, M. S. \& Brett, J. M.  1988, \pasp, 100, 1134 
\bibitem[Breger(1986)]{Breger86}  Breger, M.  1986, \apj, 309, 311 
\bibitem[Breger(1987)]{Breger87}  Breger, M.  1987, \apj, 319, 754 
\bibitem[Briggs \& Pye(2003)]{Briggs03} Briggs, K. R. \& Pye, J. P.  2003, \mnras, 345, 714 
\bibitem[Bryden et al.(2006)]{Bryden06} Bryden, G. et al. 2006, \apj, 636, 1098
\bibitem[Cambr\'{e}sy et al.(2002)]{Cambresy02} Cambr\'{e}sy, L., Beichman, C. A., 
 Jarrett, T. H., \& Cutri, R. M.  2002, \aj, 123, 2559 
\bibitem[Carpenter(2001)]{Carpenter01} Carpenter, J. M.  2001, \aj, 121, 2851 
\bibitem[Castelaz, Sellgren, \& Werner(1987)]{Castelaz87}  Castelaz, M. W., Sellgren, K., \& Werner, M. W.  
  1987, \apj, 313, 853 
\bibitem[Chen et al.(2005a)]{Chen05a} Chen, C. H., Jura, M., Gordon, K. D., \& Blaylock, M.  2005a, 
 \apj, 623, 493  
\bibitem[Chen et al.(2005b)]{Chen05b} Chen, C. H., Patten, B. M., Werner, M. W.,
  Dowell, C. D., Stapelfeldt, K. R., Song, I., Stauffer, J. R., Blaylock, M., Gordon, K. D., \& Krause, V.  
 2005b, \apj, 634, 1372
\bibitem[Daniel, Linsky \& Gagn\'{e}(2002)]{Daniel02} Daniel, K. J., Linsky, J. L., \& Gagn\'{e}, M. 
  2002, \apj, 578, 486 
\bibitem[Decin et al.(2003)]{Decin03} Decin, G., Dominik, C., Waters, L. B. F. M., \& Waelkens, C.  2003, 
 \apj, 598, 636 
\bibitem[Dermott et al.(2002)]{Dermott02} Dermott, S. F., Durda, D. D., Grogan, K., \& Kehoe, T. J. J.  2002,  
 in Asteroids III, eds. W. F. Bottke Jr., A. Cellino, P. Paolicchi,   
 \& R. P. Binzel (Tucson: Univ. Arizona Press), 423 
\bibitem[Dommanget \& Nys(2000)]{Dommanget00} Dommanget, J. \& Nys, O.  2000, \aap, 363, 991 (VizieR on-line 
  catalog I/260) 
\bibitem[Gibson \& Nordsieck(2003)]{Gibson03} Gibson, S. J. \& Nordsieck, K. H.  2003, \apj, 589, 362 
\bibitem[Gies et al.(1990)]{Gies90} Gies, D. R., McKibben, W. P., Kelton, P. W., Opal, C. B., \& Sawyer, S.   
  1990, \aj, 100, 1601 
\bibitem[Gomes et al.(2005)]{Gomes05} Gomes, R., Levison, H. F., Tsiganis, K., \& Morbidelli, A.  2005, 
 \nat, 435, 466 
\bibitem[Gordon et al.(2005)]{Gordon05} Gordon, K. D. et al.  2005, \pasp, 117, 503 
\bibitem[Gorlova et al.(2004)]{Gorlova04} Gorlova, N. et al.  2004, \apjs, 154, 448 
\bibitem[Hahn et al.(2002)]{Hahn02} Hahn, J. M., Zook, H. A., Cooper, B., \& Sunkara, B.  2002, \icarus, 158, 360 
\bibitem[Harmanec et al.(2002)]{Harmanec02} Harmanec, P., Bisikalo, D. V., Boyarchuk, A. A., \& Kuznetsov, O. A. 
   2002, \aap, 396, 937 
\bibitem[Herbig \& Simon(2001)]{Herbig01} Herbig, G. H. \& Simon, T.  2001, \aj, 121, 3138 
\bibitem[Hillenbrand(2005)]{Hillenbrand05} Hillenbrand, L. A. 2005, in  
 ``A Decade of Discovery: Planets Around Other Stars'', STScI Symposium Series 19, ed. M. Livio (astro-ph/0511083) 
\bibitem[Holland et al.(1998)]{Holland98} Holland, W. S., Greaves, J. S., Zuckerman, B., 
  Webb, R. A., McCarthy, C., Coulson, I. M., Walther, D. M., Dent, W. R. F., Gear, W. K., \& Robson, I.  1998, 
 \nat, 392, 788 
\bibitem[Jura(1990)]{Jura90} Jura, M.  1990, \apj, 365, 317 
\bibitem[Kalas et al.(2002)]{Kalas02} Kalas, P., Graham, J. R., Beckwith, S. V. W., 
  Jewitt, D. C., \& Lloyd, J. P.  2002, \apj, 567, 999 
\bibitem[Kenyon \& Bromley(2005)]{Kenyon05} Kenyon, S. J. \& Bromley, B. C. 2005, \aj, 130, 269 
\bibitem[Kharchenko, Piskunov \& Scholz(2004)]{Kharchenko04} Kharchenko, N. V., Piskunov, A. E., \& Scholz, R.-D.  2004, Astron. Nachr., 325, 439 
 (VizieR on-line catalog III/239) 
\bibitem[Kim, Zuckerman, \& Silverstone(2001)]{Kim01} Kim, S. S., Zuckerman, B., \& Silverstone M.  2001, \apj, 550, 1000 
\bibitem[Kim et al.(2005)]{Kim05} Kim, J. S. et al.  2005, \apj, 632, 659 
\bibitem[Kleine et al.(2002)]{Kleine02} Kleine, T., M\"{u}nker, C., Mezger, K., \& Palme, H.  2002, \nat, 418, 952 
\bibitem[Kraemer et al.(2003)]{Kraemer03} Kraemer, K. E., Shipman, R. F., Price, S. D., Mizuno, D. R., 
  Kuchar, T., \& Carey, S. J.  2003, \aj, 126, 1423 
\bibitem[Lada et al.(2006)]{Lada06} Lada, C. J., et al.  2006, \apj, in press (astro-ph/0511638) 
\bibitem[Lagrange, Backman, \& Artymowicz(2000)]{Lagrange00} Lagrange, A.-M., 
 Backman, D. E., \& Artymowicz, P.  2000, in Protostars 
 and Planets IV, eds. V. Mannings, A. P. Boss, \& S. S. Russell (Tucson: Univ. Arizona Press), 639 
\bibitem[Laureijs et al.(2002)]{Laureijs02} Laureijs, R. J., Jourdain de Muizon, M., Leech, K.,
 Siebenmorgen, R., Dominik, C., Habing, H. J., Trams, N., \& Kessler, M. F. 2002, \aap, 387, 285
\bibitem[Liou \& Zook(1999)]{Liou99} Liou, J.-C. \& Zook, H. A. 1999, \aj, 118, 580
\bibitem[Liu, Janes, \& Bania(1991)]{Liu91} Liu, T., Janes, K. A., \& Bania, T. M.  1991, \apj, 377, 141 
\bibitem[Low et al.(1984)]{Low84} Low, F. J. et al.  1984, \apj, 278, L19 
\bibitem[Low et al.(2005)]{Low05} Low, F. J., Smith, P. S., Werner, M., Chen, C., 
 Krause, V., Jura, M. \& Hines, D. C.  2005, \apj, 631, 1170 
\bibitem[Mart\'{i}n \& Rodr\'{i}guez(2000)]{Martin00} Mart\'{i}n, S. \& Rodr\'{i}guez, E.  2000, \aap, 358, 287 
\bibitem[Mason et al.(2001)]{Mason01} Mason, B. D., Wycoff, G. L., Hartkopf, W. I., Douglass, G. G., \& Worley, C. E.  2001, \aj, 122, 3466 
\bibitem[Mason et al.(1993)]{Mason93} Mason, B. D., Hartkopf, W. I., McAlister, H. A., \& Sowell, J. R.  1993, 
  \aj, 106, 637 
\bibitem[McAlister et al.(1989)]{McAlister89} McAlister, H. A., Hartkopf, W. I., Sowell, J. R.,  
  Dombrowski, E. G., \& Franz, O. G.  1989, \aj, 97, 510 
\bibitem[Meyer et al.(2004)]{Meyer04} Meyer, M. R. et al.  2004, \apjs, 154, 422 
\bibitem[Micela et al.(1999)]{Micela99} Micela G., Sciortino S., Harnden, F. R. Jr., Kashyap V., Rosner R., 
       Prosser C. F., Damiani F., Stauffer J., Caillault J.-P.  1999, \aap, 341, 751  
\bibitem[Moraux et al.(2003)]{Moraux03} Moraux, E., Bouvier, J., Stauffer, J. R., \& Cuillandre, J.-C. 2003, \aap, 400, 891
\bibitem[Morbidelli, Brown, \& Levison(2003)]{Morbidelli03} Morbidelli, A., Brown, M. E., \& Levison, H. F.  2003, 
  Earth, Moon, and Planets, 92, 1
\bibitem[Moro-Mart\'{i}n \& Malhotra(2002)]{MoroMartin02} Moro-Mart\'{i}n, A. \& Malhotra, R.  2002, \aj, 124, 2305 
\bibitem[Pinfield et al.(2003)]{Pinfield03} Pinfield, D. J., Dobbie, P. D., Jameson, R. F., 
  Steele, I. A., Jones, H. R. A., \& Katsiyannis, A. C.  2003, \mnras, 342, 1241 
\bibitem[Plets et al.(1997)]{Plets97} Plets, H., Waelkens, C., Oudmaijer, R. D., \& Waters, L. B. F. M.  1997, \aap, 323, 513 
\bibitem[Porter \& Rivinius(2003)]{Porter03} Porter, J. M. \& Rivinius, T.  2003, \pasp, 115, 1153 
\bibitem[Queloz et al.(1998)]{Queloz98}	Queloz, D., Allain, S., Mermilliod, J.-C., Bouvier, J., \& Mayor, M. 
  1998, \aap, 335, 183 
\bibitem[Raboud \& Mermilliod(1998)]{Raboud98} Raboud, D. \& Mermilliod, J.-C.  1998, \aap, 329, 101  
\bibitem[Rieke et al.(2004)]{Rieke04} Rieke et al.  2004, \apjs, 154, 25 
\bibitem[Rieke et al.(2005)]{Rieke05} Rieke et al.  2005, \apj, 620, 1010  
\bibitem[Siess, Dufour, \& Forestini(2000)]{Siess00} Siess, L., Dufour, E., \& Forestini, M.  2000, \aap, 358, 593 
\bibitem[Smith et al.(2006)]{Smith06}  Smith, P. S., Hines, D. C., Low, F. J., Gehrz, R. D., Polomski, E. F., 
        Woodward, C. E.  2006, \apj, in press (astro/ph 0605334)
\bibitem[Soderblom et al.(1993)]{Soderblom93} Soderblom, D. R., Stauffer, J. R., Hudon, J. D., \& Jones, B. F.  1993, \apjs, 85, 315 
\bibitem[Soderblom et al.(2005)]{Soderblom05} Soderblom, D. R., Nelan, E., Benedict, G. F.,
 McArthur, B., Ramirez, I., Spiesman, W., Jones B. F.  2005, \aj, 129, 1616
\bibitem[Spangler et al.(2001)]{Spangler01} Spangler, C., Sargent, A. I., Silverstone, M. D., Becklin, E. E., \& Zuckerman, B. 
  2001, \apj, 555, 932  
\bibitem[Su et al.(2005)]{Su05} Su, K. Y. L. et al.  2005, \apj, 628, 487 
\bibitem[Stapelfeldt et al.(2004)]{Stapelfeldt04} Stapelfeldt, K. R. et al.  2004, \apjs, 154, 458 
\bibitem[Stauffer et al.(1984)]{Stauffer84} Stauffer, J. R., Hartmann, L., Soderblom, D. R., \& Burnham, N.  1984, \apj, 280, 202
\bibitem[Stauffer \& Hartmann(1987)]{Stauffer87} Stauffer, J. R. \& Hartmann, L. W.  1987, \apj, 318, 337 
\bibitem[Stauffer et al.(2005)]{Stauffer05} Stauffer, J. R. et al.  2005, \apj, 130, 1834 
\bibitem[Strom et al.(2005)]{Strom05} Strom, R. G., Malhotra, R., Ito, T., Yoshida, F., \& Kring, D. A.  2005, Science, 309, 1847 
\bibitem[Telesco et al.(2005)]{Telesco05} Telesco, C. M., et al.  2005, \nat, 433, 133 
\bibitem[Uzpen et al.(2005)]{Uzpen05} Uzpen, B. et al.  2005, \apj, 629, 512 
\bibitem[Waters (1986)]{Waters86} Waters, L. B. F. M.  1986, \aap, 162, 121 
\bibitem[White(2003)]{White03} White, R. E.  2003, \apjs, 148, 487 
\bibitem[Young et al.(2004)]{Young04} Young, E. T. et al.  2004, \apjs, 154, 428 
\bibitem[Zuckerman \& Becklin(1993)]{Zuckerman93} Zuckerman, B. \& Becklin, E. E.  1993, \apj, 414, 793 
\bibitem[Zuckerman(2001)]{Zuckerman01} Zuckerman, B.  2001, \araa, 39, 549 
\end{thebibliography}
\end{document}